\PassOptionsToPackage{numbers,sort}{natbib}

\documentclass[a4paper,american,times,5p]{elsarticle}

\makeatletter
\def\@journal{arXiv}
\makeatother

\usepackage[T1]{fontenc}
\usepackage[utf8]{inputenc}
\usepackage{babel}

\usepackage{amsmath}
\usepackage{amsthm}
\usepackage{amssymb}
\usepackage{graphicx}
\usepackage{color}
\usepackage{bm}

\usepackage{booktabs}		% support for fancy tables
\usepackage[figuresright]{rotfloat}	% support for sideways figures and tables (always rotate counterclockwise)

\usepackage{multirow}

\usepackage{stmaryrd}
\usepackage{stackrel}

\usepackage{flushend}

\usepackage{afterpage}

\makeatletter

\providecommand{\tabularnewline}{\\}
\newcommand{\lyxdot}{.}

\definecolor{note_fontcolor}{rgb}{1, 0.667969, 0}

\usepackage{environ}
\usepackage[colorinlistoftodos,prependcaption,textsize=small]{todonotes}
\RenewEnviron{lyxgreyedout}{\todo[inline]{\protect\BODY}}
\usepackage{etoolbox}

\makeatletter
\@ifundefined{showcaptionsetup}{}{%
 \PassOptionsToPackage{caption=false}{subfig}}
\usepackage{subfig}
\makeatother

  \theoremstyle{definition}
  
  \newtheorem*{defn*}{\protect\definitionname}
  \providecommand{\definitionname}{Definition}
  \theoremstyle{plain}
  
  \providecommand{\lemmaname}{Lemma}
  \theoremstyle{remark}
  \newtheorem*{rem*}{\protect\remarkname}
  \providecommand{\remarkname}{Remark}
  \theoremstyle{plain}
  
  \providecommand{\theoremname}{Theorem}

\renewcommand{\vec}[1]{\bm{#1}}

\usepackage{hyperref}

\makeatother

\newcommand{\R}{\mathbb{R}}

\newcommand{\J}{\mathcal{J}}

\newcommand{\NN}[1]{\left\{  1,\dots,#1\right\}  }

\begin{document}
\begin{frontmatter}

\title{Three-dimensional phase-field simulations of water freezing and thawing at pore-scale}

%\titlerunning{....}

\author[fnspe]{Pavel Strachota\corref{cor1}}\ead{pavel.strachota@fjfi.cvut.cz}

\cortext[cor1]{Corresponding author. Phone: +420 778 546 112}

\address[fnspe]{Department of Mathematics, Faculty of Nuclear Sciences
and Physical Engineering, Czech Technical University in Prague. Trojanova
13, 120 00 Praha 2, Czech Republic}

\begin{abstract}
This work deals with numerical simulation of water freezing and thawing
in a complex three-dimensional geometry of a porous medium. The porous
structure is represented by a virtual container filled with glass
beads. Phase transition modeling is approached at both macro-scale
and micro-scale, combining heat transfer in a heterogeneous medium
and a phase-field approximation of the Gibbs-Thomson relation by means
of the Allen-Cahn equation. The formulation of the model contains
novel components tailored for the given purpose. At the macro-scale,
surface tension effects are negligible and phase transition focusing
based on temperature can replace the Allen-Cahn equation. In contrast
to that, simulations of equilibrium states at the micro-scale allow
to eliminate the heat equation by assuming constant supercooling.
For numerical solution, an efficient hybrid parallel algorithm based
on the finite volume method and the Runge-Kutta-Merson solver with
adaptive time stepping are employed. The results of different model
variants at different scales are discussed. In a parametric study,
the full phase-field model is demonstrated to deliver consistent results
across a wide range of surface tension values, exhibiting curvature-induced
premelting if surface tension is artificially exaggerated. As surface
tension tends to the realistic values, the results of the phase-field
approach those of the simplifed temperature-driven phase transition
model. In addition, micro-scale simulations of water freezing at different
supercooling values aim to predict the unfrozen water content and
compare the results with data from literature. Numerical stability,
accuracy, and computational costs are also discussed.
\end{abstract}


%%%%% AMS/MSC/PACs/Keywords %%%%%%%%%%%
%% AMS 2020 Subject Classification codes legend:
%% https://mathscinet.ams.org/msc/msc2020.html
%% ------------------------------------------------------------------------
%% 35K51 Partial differential equations | Parabolic equations and systems |
%%       Initial-boundary value problems for second-order parabolic systems
%% 35K57 Partial differential equations | Parabolic equations and systems |
%%       Reaction-diffusion equations
%%
%% 80A22 Classical thermodynamics, heat transfer | Stefan problems, phase changes, etc.
%% 74N05 Mechanics of deformable solids | Crystals in solids
%% ------------------------------------------------------------------------

\end{frontmatter}

% set up the base length for image width adjustment
% use as:  \includegraphics[width=0.9\columnwidth] in LyX
% (occurrences of \columnwidth in the LaTeX file exported from LyX are replaced by \figwidth automatically by the LyX-import.sh script)
\newlength \figwidth
\setlength \figwidth {1.0\columnwidth}

%% --------------------- THE PAPER TEXT ITSELF ---------------------
%LATEX_EXPORT_BEGIN

%\setuptodonotes{disable}

\section{Introduction}

Scientific investigation of freezing and thawing of water in porous
media has been of continuous interest. Understanding these phenomena
and their implications on the mechanics of materials (such as frost
heave or cracking) is particularly important in environmental and
civil engineering applications in cold regions. It has been observed
that the mechanical effects of freezing and thawing in soils and other
porous structures are influenced by the presence of some amount of
unfrozen water even at subzero temperatures. This phenomenon generally
known as premelting \cite{Dash-1995-premelting} is caused by several
mechanisms related to the presence of solid surfaces in the complex
geometry of the pores. Experimental methods for measuring the unfrozen
water content are available \cite{Yoshikawa-2005-unfrozen_water_content_measurement,Zhou_Zhou-Unfrozen_content_measurement-2013,Watanabe-2017-unfrozen_water_measurements}
and a number of mathematical models for its prediction have been proposed
\cite{Cahn_Dash_Fu-Premelting_monosized_powders-1992,Zhou_Zhou-Generalized-Clapeyron-unfrozen-2018,Bi-2023-Unfrozen_water-FEarthS,Xusheng-2023-unfrozen-water-theoretical}.
Just recently, imaging of the premelting processes with atomic resolution
has been successfully carried out \cite{Nature-Premelting-atomic_resolution}.

One of the causes of the presence of liquid water at temperatures
below zero is the surface tension at the liquid--ice interface. As
the ice crystal penetrates the narrow apertures of the pores, it develops
regions with high positive curvature where the surface tension opposes
further crystal growth thanks to the Gibbs-Thomson effect. This is
known as curvature-induced premelting \cite{Rempel_Wettlaufer-Premelting_dynamics-2004,Dash-2006-premelting}.
In addition, a host of complex phenomena at contact of the solid material
and water (or other substance subject to phase transitions) leads
to the creation of a very thin liquid film adjacent to the solid particles'
surface, which is called interfacial (surface) premelting \cite{Wettlaufer_Worster-Premelting_review_2006,Dash-2006-premelting}.
In this type of premelting, the thermodynamic equilibrium, the existence
of impurities and chemical processes at the particle's surface come
into play.

In addition to the theoretical and experimental methods cited above,
relevant information on the presence of liquid water might be obtained
from detailed numerical simulations of freezing and thawing with pore-scale
resolution. As a powerful tool for this purpose, the phase-field methods
\cite{Provatas_Elder-PF_Book,Boettinger-SolidificationMicro_Overview}
can been used. These methods originally developed for describing the
evolution of solid material microstructure \cite{free-energy-cahn-hil,Allen-Cahn-orig}
have seen successful developments over the past decades, as numerical
implementations targeted simulations of solidification of pure materials
\cite{Wheeler-PF-Dendrites,Kobayashi-PF-Dendritic,Kupferman-Num_Study_Morphological_Diagram,Elliott_Gardiner-DoubleObstacle_PF}
and alloys \cite{Wheeler-PF-Binary_Alloys,Nestler-PhaseField_Multicomponent}
in two and later three spatial dimensions \cite{Karma_Rappel-Num_sim_3D_dendrites}.
More recently, high performance GPU-based solvers have been utilized
\cite{Rojas_Takaki_Ohno-PF-LBM-dendrite_convection,Takaki-GPU_PF_LBM-growth_and_motion}
for large scale 3D crystal growth simulations and applications in
many other domains have been found \cite{PF-NS-analysis-fluid-interaction,PF-fracture-crack-propagation,PF-Asy-Analysis,PF-multi-comp-flow}.

As for modeling freezing and thawing in porous media, phase-field
models coupled with poro-thermo-hydro-mechanics \cite{Lu-2011-PF-freezing-porous-theoretical}
have been proposed and recent works such as \cite{Zak-Benes-2018-ISPMA14,Zak-Benes-2023-CMAME}
or \cite{Sweidant-2020-Coupled-PF-model-frost-porous,Yang-2023-Freezing-at-pore-scale-numerical_experimental}
present numerical simulations with resolved pore matrix in two spatial
dimensions. To the author's best knowledge, however, three-dimensional
simulations at pore scale such as \cite{Gharedaghloo-2020-Pore-scale-3D-simulations}
have been rare.

In this work, we follow up on our previous results in phase-field
crystal growth modeling \cite{PF-Focusing-Latent-Heat} and propose
a basic model of water freezing and thawing in a three-dimensional
domain containing a porous medium. At this stage, only the heat transfer
and the Gibbs-Thomson effect govern the evolution of the phase transitions.
Mechanical interactions, fluid flow, and the complex mechanisms of
interfacial premelting are currently not included.

The model equations are laid out in Section \ref{sec:mathematical-model},
which discusses the full model and its special variants that are suited
for macro-scale and micro-scale situations, respectively. The problem
formulation including the initial and boundary conditions is loosely
inspired by the setup of the experiments described in \cite{Sklenar-DP-MRI-freezing}
and \cite{Snehota-et_al-MRI-freezing}, where the dynamic freezing
and thawing phenomena in porous media were investigated by means of
magnetic resonance imaging (MRI). Numerical solution of the problem
is based on our previous works \cite{ALGORITMY2016}, \cite{ISPMA14-Pavel_Ales-ActaPhPoloA},
\cite{ALGORITMY2020-Hybrid_Parallel_Polycrystalline} and is briefly
described in Section \ref{subsec:Numerical-solution}.

In Section \ref{sec:Results-freezing-thawing}, results of two types
of numerical simulations are discussed. In Sections \ref{subsec:Parameters-setup}-\ref{subsec:Macro-scale-results},
the simulation cases represent an analogue of the above mentioned
experiment. The results are discussed with the aim to assess the qualitative
behavior of the proposed models and with the prospect of future validation
against measurement data. In Section \ref{subsec:Micro-scale-results},
a scaled down version of the problem is used to estimate the unfrozen
water content dependence on supercooling and comparison with the results
from literature \cite{Cahn_Dash_Fu-Premelting_monosized_powders-1992,Maruyama_Dash-Interfacial_melting_graphite_talc-1992,Dash-2006-premelting}
is provided. The influence of mesh resolution and computational costs
are also discussed.

\section{\protect\label{sec:mathematical-model}Mathematical models of heat
transfer and phase transitions in a heterogeneous medium}

Let us lay out three similar mathematical models of heat transfer,
freezing, and thawing formulated for the following situation:
\begin{itemize}
\item The computational domain $\Omega=\left(0,L_{1}\right)\times\left(0,L_{3}\right)\times\left(0,L_{3}\right)\subset\R^{3}$
represents the interior of a cuboidal container partially filled by
solid spheres (glass beads) forming a porous medium.
\item A thin glass cap is placed inside $\Omega$, next to its top boundary
face.
\item All remaining void space is fully saturated by water.
\item The boundary of the domain can be split in two parts $\partial\Omega=\Gamma_{\text{cap}}\cup\Gamma_{\text{wall}}$,
where
\begin{itemize}
\item $\Gamma_{\text{cap}}$ represents the top face of $\Omega$ immediately
adjacent to the glass cap, which is uniformly cooled or heated to
the prescribed temperature, and
\item $\Gamma_{\text{wall }}$ represents the side walls and the bottom
of the container, where ideal thermal insulation is applied.
\end{itemize}
\item The time interval $\J=\left(0,t_{\text{freeze}}+t_{\text{thaw}}\right)$
relevant for the modeling consists of a freezing phase for $t\in\left(0,t_{\text{freeze}}\right)$
when $\Gamma_{\text{cap}}$ is cooled, immediately followed by a thawing
phase for $t\in\left[t_{\text{freeze}},t_{\text{freeze}}+t_{\text{thaw}}\right)$
when $\Gamma_{\text{cap}}$ is heated.
\end{itemize}
In the course of the following explanation, we often refer to our
previous work \cite{PF-Focusing-Latent-Heat} which can provide deeper
understanding and theoretical background for the discussed topics.

\subsection{\protect\label{subsec:General-concept}General concept and the involved
physical quantities}

In accordance with the intended application, the models only consider
heat conduction, phase transitions associated with latent heat release
or consumption, and curvature-related phase interface dynamics. Mechanical
interactions and fluid flow are currently not included. All models
resolve the evolution of the temperature field $T:\bar{\J}\times\bar{\Omega}\to\R$.
Their difference consists in the treatment of the phase transitions
of water by means of a scalar order parameter $\phi:\J\times\bar{\Omega}\to\left[0,1\right]$.
$\phi$ is a smooth function indicating the presence of liquid water
if $\phi\left(t,\vec{x}\right)\approx0$ or ice if $\phi\left(t,\vec{x}\right)\approx1$.
Between the solid and liquid regions, a smooth transition is formed,
implicitly representing a diffuse phase interface as
\begin{equation}
\Gamma\left(t\right)=\left\{ \left.\vec{x}\in\Omega\vphantom{\frac{1}{2}}\right|\phi\left(t,\vec{x}\right)=\frac{1}{2}\right\} .\label{eq:phase-interface}
\end{equation}

\begin{figure}
\begin{centering}
\includegraphics[width=0.98\figwidth]{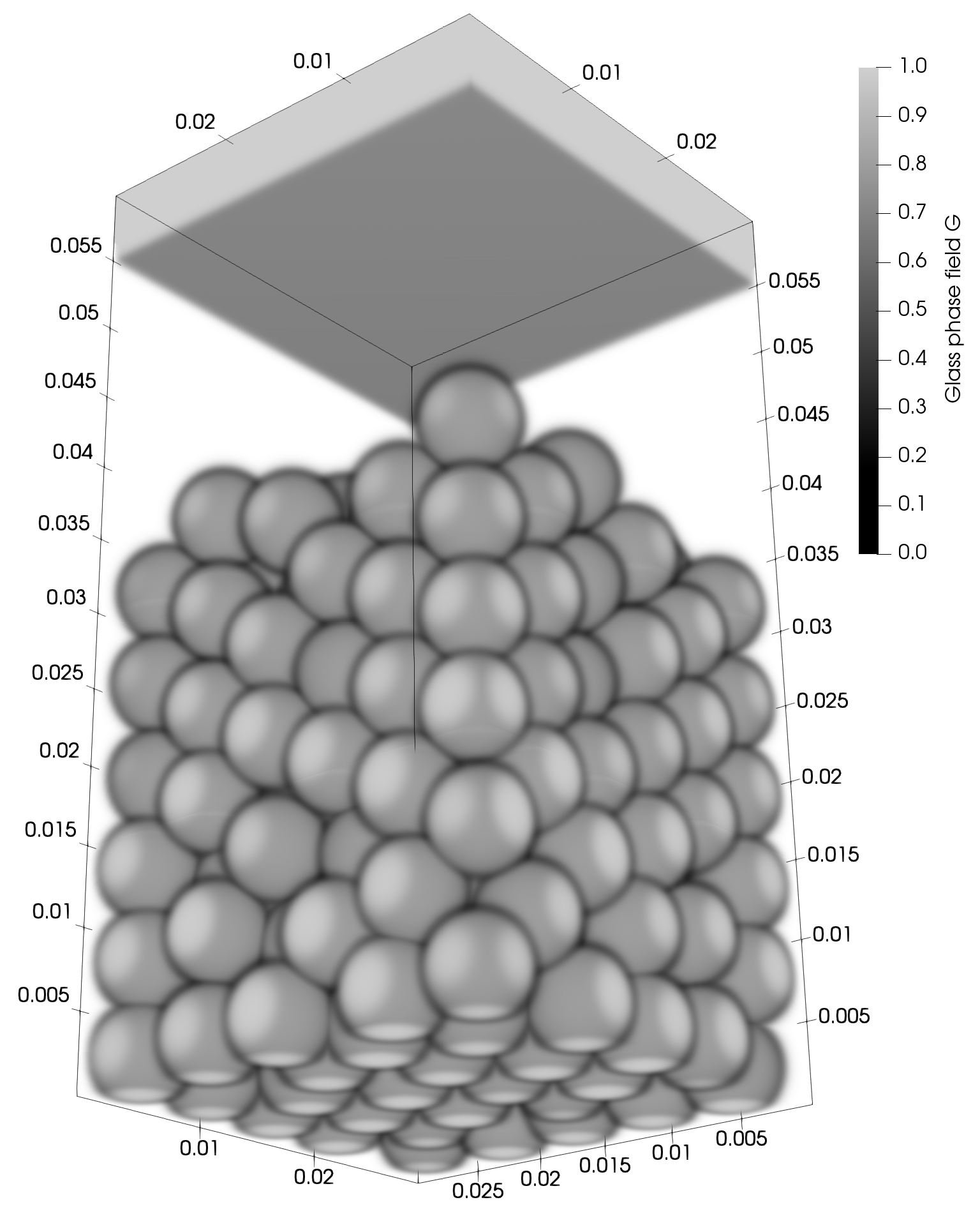}
\par\end{centering}
\caption{\protect\label{fig:spheres-and-cap}Geometrical configuration of the
glass within the container $\Omega$ used for simulations in Section
\ref{sec:Results-freezing-thawing}. The vertical axis is $x_{3}$.
More details are given in Table \ref{tab:simulation-parameters-setup}.}
\end{figure}

\begin{itemize}
\item In the phase-field models (Section \ref{subsec:Phase-Field-Approach}),
$\phi$ is an unknown variable (referred to as the phase field) and
its evolution is governed by the Allen-Cahn equation \cite{Allen-Cahn-orig},
which allows for the approximation of the Gibbs-Thomson effect caused
by surface tension at the solid--liquid interface. The thickness
of the diffuse interface (i.e., of the region where $\phi$ differs
substantially from both $0$ and $1$) can be controlled by a scalar
parameter $\xi>0$ \cite{PF-Focusing-Latent-Heat}.
\item In the simplified model (Section \ref{subsec:Macro-scale-approximation}),
the surface tension is neglected. The primary unknown is the temperature
field $T$ and $\phi$ can be expressed as a function of $T$.
\end{itemize}
The presence of glass within $\Omega\subset\R^{3}$ is modeled using
a scalar indicator function $G\left(\vec{x}\right)$ independent of
time, where, in principle
\[
G\left(\vec{x}\right)=\begin{cases}
0 & \text{if water is located at }\vec{x},\\
1 & \text{if glass is located at }\vec{x}.
\end{cases}
\]
For well-posedness of the model and for numerical stability, smooth
transitions between glass and water are employed. Given $n$ spheres
with identical radius $r$ centered at $\vec{x}_{i}$, $i\in\NN n$,
the glass field representing the beads is calculated as
\[
G_{\text{balls}}\left(\vec{x}\right)=\max_{i\in\NN n}\frac{1}{2}\left(1-\tanh\left(\frac{1}{2\xi_{G}}\left(\left|\vec{x}-\vec{x}_{i}\right|-r\right)\right)\right).
\]
The final form of the field $G$ is obtained by adding the glass cap
as
\[
G\left(\vec{x}\right)=\max\left\{ G_{\text{balls}}\left(\vec{x}\right),\frac{1}{2}\left(1+\tanh\left(\frac{1}{2\xi_{G}}\left(x_{3}-x_{3.\text{cap}}\right)\right)\right)\right\} ,
\]
where $x_{3,\text{cap}}$ is the $x_{3}$ coordinate of the bottom
of the glass cap and $\xi_{G}>0$ plays a similar role in shaping
$G$ as $\xi$ does in shaping $\phi$. The resulting geometrical
configuration used for the simulations later in Section \ref{sec:Results-freezing-thawing}
is depicted in Figure \ref{fig:spheres-and-cap}. The positions of
the particle centers were obtained by running a simulation of particle
settling using the discrete element method \cite{ALGORITMY2024-DEM_simulations-preprint}.

Even though dimensionless formulations are common in the field \cite{PF-Focusing-Latent-Heat,Benes-Asymptotics,Benes-Math_comp_aspects_solid,CMA2023-PF-Optimization-numerical,JDCS2023-PF-OptimalControl-Theory},
it is beneficial for our application to formulate all equations and
relations in SI units. All physical parameters and their respective
units are summarized in Table \ref{tab:Physical-quantities}. In order
to evaluate the heat capacity $c$, the thermal conductivity $\lambda$,
and the mass density $\rho$ at any point $\vec{x}\in\Omega$, linear
interpolation based on the values of the glass field $G$ and the
phase field $\phi$ is employed. Let $w_{\text{G}},w_{\text{L}},w_{\text{I}}$
stand for the values of the quantity $w\in\left\{ c,\lambda,\rho\right\} $
for glass (G), liquid water (L) and ice (I), respectively. Then the
value of $w$ at the given time $t$ and point $\vec{x}\in\Omega$
is calculated by the blending formula
\begin{equation}
w\left(t,\vec{x}\right)=G\left(\vec{x}\right)w_{\text{G}}+\left(1-G\left(\vec{x}\right)\right)\left[\phi\left(t,\vec{x}\right)w_{\text{I}}+\left(1-\phi\left(t,\vec{x}\right)\right)w_{\text{L}}\right].\label{eq:material-properties-blending-formula}
\end{equation}
\begin{table*}
\caption{\protect\label{tab:Physical-quantities}Physical properties of the
individual materials used in the simulations in Section \ref{sec:Results-freezing-thawing}.}

\centering{}%
\begin{tabular}{cclccc}
\toprule
Quantity & SI Unit & Description & \multicolumn{3}{c}{Value}\tabularnewline
 &  &  & Liquid water (L) & Ice (I) & Glass (G)\tabularnewline
\midrule
$\rho$ & $\text{kg}\,\text{m}^{-3}$ & density & 997 & 917 & 2500\tabularnewline
$c$ & $\text{J}\,\text{kg}^{-1}\text{K}^{-1}$ & specific heat capacity & 4180 & 2050 & 840\tabularnewline
$\lambda$ & $\text{W}\,\text{m}^{-1}\,\text{K}^{-1}$ & heat conductivity & 0.6 & 2.22 & 1.1\tabularnewline
$L$ & $\text{J}\,\text{kg}^{-1}$ & specific latent heat of fusion of water & \multicolumn{2}{c}{$3.34\cdot10^{5}$} & \tabularnewline
$T^{*}$ & $\text{K}$ & freezing/melting point of bulk water & \multicolumn{2}{c}{273.15} & \tabularnewline
$\sigma$ & $\text{J}\,\text{m}^{-2}$ & surface tension at the ice--liquid interface (see \cite{Ketcham_Hobbs1969-surface_energy_ice}) & \multicolumn{2}{c}{0.033} & \tabularnewline
$\Delta s$ & $\text{J}\,\text{m}^{-3}\,\text{K}^{-1}$ & entropy difference per unit volume $\Delta s=\frac{\rho_{\text{I}}L}{T^{*}}$ & \multicolumn{2}{c}{$1.121\cdot10^{6}$} & \tabularnewline
$\mu$ & $\text{m}\,\text{s}^{-1}\,\text{K}^{-1}$ & ice--liquid interface mobility (see \cite{Shibkov-interface_mobility,Wheeler-PF-Binary_Alloys,Wheeler-PF-Dendrites}) & \multicolumn{2}{c}{$10^{-4}$} & \tabularnewline
$\alpha$ & $\text{m}^{-2}\,\text{s}$ & coef. of attachment kinetics $\alpha=\frac{\Delta s}{\mu\sigma}$ & \multicolumn{2}{c}{$3.39782\cdot10^{11}$} & \tabularnewline
\bottomrule
\end{tabular}
\end{table*}

\subsection{\protect\label{subsec:Phase-Field-Approach}Phase-field approach}

We begin with the formulation of the full phase-field model with
surface tension, consisting of the heat equation and the Allen-Cahn
equation in dimensional form. The simplified model without surface
tension will be derived later in Section \ref{subsec:Macro-scale-approximation}.
Based on the results of our previous work \cite{PF-Focusing-Latent-Heat}
and taking the glass field $G$ into account, we propose the governing
equations in $\J\times\Omega$ in the form
\begin{align}
\rho c\frac{\partial T}{\partial t} & =\nabla\cdot\left(\lambda\nabla T\right)+\rho L\frac{\partial\phi}{\partial t},\label{eq:heat-equation}\\
\alpha\xi^{2}\frac{\partial\phi}{\partial t} & =I_{\text{W}}\left(G\right)\left[\xi^{2}\nabla^{2}\phi+f\left(T,\phi,\nabla\phi;\xi\right)\right].\label{eq:Allen-Cahn-equation}
\end{align}
The water indicator function
\begin{equation}
I_{\text{W}}\left(G\right)=\text{max}\left(0,1-\zeta G\right)\label{eq:water-indicator}
\end{equation}
for some $\zeta\geq1$ ensures that $\phi$ can only change in the
region occupied by water, as discussed later in Section \ref{subsec:Parameters-setup}.
The reaction term $f$ on the right hand side of (\ref{eq:Allen-Cahn-equation})
generally consists of the derivative of the double-well potential
\cite{Caginalp-Stefan_HeleShaw_PF} and a term related to latent heat
release or consumption. For the model to remain valid without restrictions
on supercooling, latent heat exchange must be focused at the diffuse
interface. To achieve this, we employ gradient-based focusing \cite{Benes-Comp_studies_Diff_Interface,PF-Focusing-Latent-Heat,Benes-Math_comp_aspects_solid}
(further referred to as the \textbf{GradP} model) by using
\begin{equation}
f\left(T,\phi,\nabla\phi;\xi\right)=2\phi\left(1-\phi\right)\left(\phi-\frac{1}{2}\right)+\xi^{2}\mu\alpha\left|\nabla\phi\right|\left(T^{*}-T\right)\label{eq:reaction-term-GradP}
\end{equation}
or the \textbf{$\mathbf{\Sigma}$P1-P} model recently proposed in
our work \cite{PF-Focusing-Latent-Heat}{\small
\begin{align}
 & f\left(T,\phi,\nabla\phi;\xi\right)\label{eq:reaction-term-SigmaP1-P}\\
= & 2\phi\left(1-\phi\right)\left(\phi-\frac{1}{2}+\frac{1}{2}\xi b\left(\varepsilon_{0},\varepsilon_{1}\right)\mu\alpha\Sigma\left(\phi;\varepsilon_{0},\varepsilon_{1}\right)\Sigma\left(1-\phi;\varepsilon_{0},\varepsilon_{1}\right)\left(T^{*}-T\right)\right).\nonumber
\end{align}
}In (\ref{eq:reaction-term-SigmaP1-P}), $\Sigma\left(w;\varepsilon_{0},\varepsilon_{1}\right)$
is defined as a differentiable sigmoid function in the form
\begin{equation}
\Sigma\left(w;\varepsilon_{0},\varepsilon_{1}\right)=\begin{cases}
0 & w\leq\varepsilon_{0},\\
1 & w\geq\varepsilon_{1},\\
\frac{3\left(w-\varepsilon_{0}\right)^{2}}{\left(\varepsilon_{1}-\varepsilon_{0}\right)^{2}}-\frac{2\left(w-\varepsilon_{0}\right)^{3}}{\left(\varepsilon_{1}-\varepsilon_{0}\right)^{3}} & w\in\left(\varepsilon_{0},\varepsilon_{1}\right).
\end{cases}\label{eq:Sigma-limiter}
\end{equation}
with a shape controlled by the parameters $\varepsilon_{0},\varepsilon_{1}$.
Note that in contrast to \cite{PF-Focusing-Latent-Heat}, the $\Sigma$-limiter
(\ref{eq:Sigma-limiter}) is used twice in (\ref{eq:reaction-term-SigmaP1-P})
to avoid phase transitions far from the interface during both freezing
and thawing. The coefficient of attachment kinetics $\alpha$ is bound
to surface tension $\sigma$ by the relation
\begin{equation}
\alpha=\frac{\Delta s}{\mu\sigma}.\label{eq:attachment-kinetics-coeff}
\end{equation}

The evolution of the ice surface $\Gamma\left(t\right)$ defined by
(\ref{eq:phase-interface}) and governed by the phase-field model
(\ref{eq:heat-equation})-(\ref{eq:attachment-kinetics-coeff}) approximates
the Gibbs-Thomson law
\begin{equation}
\mu\alpha\left(T^{*}-T\right)=\kappa_{\Gamma}+\alpha v_{\Gamma},\label{eq:Gibbs-Thomson}
\end{equation}
where $\kappa_{\Gamma}$ is the mean curvature of $\Gamma\left(t\right)$
and $v_{\Gamma}$ is its normal velocity in the direction out of the
solid (ice) subdomain. More precisely, (\ref{eq:Gibbs-Thomson}) is
recovered asymptotically as the diffuse interface thickness $\xi$
tends to zero, provided that the factor $b\left(\varepsilon_{0},\varepsilon_{1}\right)$
in (\ref{eq:reaction-term-SigmaP1-P}) is calculated accordingly \cite[Sect. 3.1]{PF-Focusing-Latent-Heat}.
If $\varepsilon_{0},\varepsilon_{1}\ll1$, then $b\left(\varepsilon_{0},\varepsilon_{1}\right)\approx1$.
However, we use $b\left(\varepsilon_{0},\varepsilon_{1}\right)=1$
in this work as it has been found in \cite{PF-Focusing-Latent-Heat}
that the difference in the results is not significant.

Plugging $v_{\Gamma}=0$ into (\ref{eq:Gibbs-Thomson}), it is obvious
that the equilibrium temperature $T$ at interfaces with positive
mean curvature is below the melting point $T^{*}$, which is one of
the causes of freezing point depression in porous media \cite{Rempel_Wettlaufer-Premelting_dynamics-2004}.
We try to verify this numerically by using the above phase-field models.

\subsubsection*{Initial and boundary conditions}

In accordance with the situation described at the beginning of Section
\ref{sec:mathematical-model}, the problem formulation based on equations
(\ref{eq:heat-equation})-(\ref{eq:attachment-kinetics-coeff}) is
closed by the initial conditions
\begin{align}
\left.T\right|_{t=0} & =T_{\text{ini}},\label{eq:init-cond-temperature}\\
\left.\phi\right|_{t=0} & =\phi_{\text{ini}},\label{eq:init-cond-PF}
\end{align}
and boundary conditions
\begin{align}
\left.T\right|_{\Gamma_{\text{cap}}} & =\begin{cases}
T_{\text{freeze}} & t\in\left(0,t_{\text{freeze}}\right),\\
T_{\text{thaw}} & t\in\left(t_{\text{freeze}},t_{\text{freeze}}+t_{\text{thaw}}\right).
\end{cases}\label{eq:BC-temperature-cap}\\
\left.\nabla T\cdot\vec{n}\right|_{\Gamma_{\text{wall}}} & =0\qquad t\in\J,\label{eq:BC-temperature-wall}\\
\left.\nabla\phi\cdot\vec{n}\right|_{\partial\Omega} & =0\qquad t\in\J.\label{eq:BC-PF}
\end{align}
As explained in \cite{PF-Focusing-Latent-Heat}, both the \textbf{GradP}
and the \textbf{$\mathbf{\Sigma}$P1-P} phase-field models focus the
phase transitions and latent heat interchange to an existing phase
interface. It is therefore necessary to initiate the freezing process
at the surface of a nucleation site, which is modeled as a region
$\Omega_{\text{I,ini}}=\left\{ \left.\vec{x}\in\Omega\vphantom{\frac{1}{2}}\right|\phi\left(0,\vec{x}\right)>\frac{1}{2}\right\} $
defined by the initial condition (\ref{eq:init-cond-PF}).

\subsection{\protect\label{subsec:Macro-scale-approximation}Macro-scale approximation
with phase transition focusing by temperature}

For studying the freezing and thawing processes with negligible surface
tension, i.e., the limit case $\sigma\to0$, $\alpha\to+\infty$,
the phase-field models introduced in Section \ref{subsec:Phase-Field-Approach}
cannot be used. They rely on the existence of a diffuse phase interface
emerging as a consequence of the reaction--diffusion nature of equation
(\ref{eq:Allen-Cahn-equation}).

When the surface tension is neglected, let us assume that the phase
field $\phi$ can be related solely to the value of temperature. To
this end, we propose a smooth dependence

\begin{equation}
\phi=\varphi\left(T\right)=\frac{1}{2}I_{\text{W}}\left(G\right)\left(1-\tanh\left(\gamma\left(T-T^{*}\right)\right)\right),\label{eq:explicit-p-dependence-on-T}
\end{equation}
where $\gamma$ is a material-specific parameter controlling the rate
of the phase transition as the temperature passes $T^{*}$. By plugging
(\ref{eq:explicit-p-dependence-on-T}) into (\ref{eq:heat-equation}),
a single governing equation arises in the form
\begin{equation}
\rho\left[c-L\varphi'\left(T\right)\right]\frac{\partial T}{\partial t}=\nabla\cdot\left(\lambda\nabla T\right).\label{eq:combined-heat-xfer-phase-transition}
\end{equation}
Note that by differentiating (\ref{eq:explicit-p-dependence-on-T})
with respect to time, we get the formal replacement of the Allen-Cahn
equation (\ref{eq:Allen-Cahn-equation}) simply as
\[
\frac{\partial\phi}{\partial t}=\varphi'\left(T\right)\frac{\partial T}{\partial t}.
\]

Equations (\ref{eq:explicit-p-dependence-on-T})--(\ref{eq:combined-heat-xfer-phase-transition})
with the initial condition (\ref{eq:init-cond-temperature}) and the
boundary conditions (\ref{eq:BC-temperature-cap})--(\ref{eq:BC-temperature-wall})
will be referred to as the \textbf{Temp} model.

\subsection{\protect\label{subsec:Micro-scale-approximation}Micro-scale approximation
with constant temperature}

With the freezing front velocity given (approximately) by the Gibbs-Thomson
relation (\ref{eq:Gibbs-Thomson}), the crystal size grows or shrinks
(roughly) in proportion to time. To see the crystal evolution at the
spatial scale reduced by the scaling factor $S\ll1$, it is natural
to scale the time variable by the same factor. The scale reduction
(including the scaling of the phase interface thickness parameter
$\xi$) causes the solution of the model (\ref{eq:heat-equation})-(\ref{eq:attachment-kinetics-coeff})
to behave like on the original scale, but with the heat conductivity
$\lambda$ and the surface tension $\sigma$ \emph{increased} by the
factor $1/S$. On the micro-scale, the heat transfer becomes so fast
that the temperature can be considered constant in space and the heat
equation (\ref{eq:heat-equation}) can be effectively eliminated from
the model.

In order to investigate freezing under a constant supercooling, this
approximation corresponds to choosing $\lambda=0,L=0$, and $t_{\text{thaw}}=0$
in the full model introduced in Section \ref{subsec:Phase-Field-Approach}.

\subsection{\protect\label{subsec:Numerical-solution}Numerical solution}

The numerical solution is based on the method of lines \cite{Method-of-lines-book},
employing a finite volume scheme with second order flux approximation
for spatial discretization \cite{ENUMATH2011} and the 4th order Runge-Kutta-Merson
integrator with adaptive time stepping \cite{Christiansen-RK-Merson}.
The finite volume mesh is uniform, dividing the cuboidal domain $\Omega$
into a grid of $N_{1}\times N_{2}\times N_{3}$ rectangular cells.
The solver takes advantage of our efficient hybrid OpenMP \cite{OpenMP_Dagum,OpenMP4_0}
/ MPI \cite{MPI-book-3_1} parallel implementation introduced in \cite{ALGORITMY2016},
using a one-dimensional domain decomposition into rectangular blocks
along the $x_{3}$ axis and multithreaded processing of each block
on a multicore compute node. Thanks to this approach, the mesh resolution
can be chosen high enough to fully resolve the pore matrix and investigate
the geometry of liquid--solid interface within the pores.

\section{\protect\label{sec:Results-freezing-thawing}Simulations of freezing
and thawing}

Let us present and compare the simulation results using the models
introduced in Sections \ref{subsec:Phase-Field-Approach}-\ref{subsec:Micro-scale-approximation}.
In general, the aim is to verify the viability of three-dimensional
simulations of freezing and thawing at pore scale before the planned
validation against experimental results and further model improvements.

In Section \ref{subsec:Parameters-setup}, the setting of the complete
set of parameters for macro-scale simulations of freezing and thawing
dynamics inspired by the experiment are explained and justified. In
this scenario, the size of the pores is so large that the Gibbs-Thomson
effect plays a negligible role. However, to assess its influence on
the qualitative behavior of freezing and thawing and to verify the
numerical properties of the models, the surface tension of water is
artificially exaggerated and a parametric study is performed. The
results are summarized in Section \ref{subsec:Macro-scale-results}.

Afterward in Section \ref{subsec:Micro-scale-results}, the complete
geometry is scaled down to simulate equilibrium situations at micro-scale,
where the real surface tension of water has significant effects. In
this situation, curvature-induced premelting contributes to the presence
of unfrozen water. Its content is calculated from simulations with
different values of constant supercooling and the results are compared
with data from literature \cite{Cahn_Dash_Fu-Premelting_monosized_powders-1992,Maruyama_Dash-Interfacial_melting_graphite_talc-1992,Dash-2006-premelting}.

\subsection{\protect\label{subsec:Parameters-setup}Parameters setup for macro-scale
simulations}

The simulation cases of the three models (\textbf{GradP}, \textbf{$\mathbf{\Sigma}$P1-P},
and \textbf{Temp}) involve a set of parameters summarized in Table
\ref{tab:simulation-parameters-setup}. The vessel dimensions of $3\times3\times6\text{ cm}$,
the duration of the experiment ($t_{\text{freeze}}=5\text{ hours}$
of cooling followed by $t_{\text{thaw}}=5\text{ hours}$ of heating),
and the corresponding settings of the temperatures $T_{\text{freeze}}=-25\,^{\circ}\text{C}$,
$T_{\text{thaw}}=+20\,^{\circ}\text{C}$ are inspired by the setup
of the experiment \cite{Sklenar-DP-MRI-freezing,Snehota-et_al-MRI-freezing}.
The porous geometry inside the vessel $\Omega$ is described in Section
\ref{sec:mathematical-model}. The physical properties of water and
a representative glass type are given in Table \ref{tab:Physical-quantities}.
The setting of some of the other parameters deserves further explanation,
which is given below.

\subsubsection*{Surface tension, mobility, and attachment kinetics coefficient}

The surface tension $\sigma$ is expressed in the phase-field models
in terms of the attachment kinetics coefficient $\alpha$ by the relationship
(\ref{eq:attachment-kinetics-coeff}). However, the realistic value
of $\sigma$ is too small (and $\alpha$ too large, see Table \ref{tab:Physical-quantities})
for the full phase field model to work at this scale. In other words,
the diffusion strength in the Allen-Cahn equation (\ref{eq:Allen-Cahn-equation})
is insufficient for the diffuse phase interface to be formed. To compare
the behavior of the models as the surface tension decreases and find
their operating range, three different and much exaggerated values
of $\sigma$ were considered in the simulations, as summarized in
Table \ref{tab:simulation-parameters-setup}:
\begin{itemize}
\item The value of $\alpha$ corresponding to ``large'' $\sigma$ is chosen
as $\alpha=\rho_{L}c_{L}$. With this setting, the thermal diffusion
in equation (\ref{eq:heat-equation}) and the phase field diffusion
in equation (\ref{eq:Allen-Cahn-equation}) occur at a comparable
rate, which is favorable for the time stepping in the numerical solution.
\item The value of $\alpha$ corresponding to ``small'' $\sigma$ is three
times larger (so $\sigma$ is three times smaller).
\item The value of $\alpha$ corresponding to ``tiny'' $\sigma$ is chosen
such that $\sigma$ is still approximately $1000\times$ larger than
the realistic value.
\end{itemize}
For the purposes of our simulations, the value of the mobility $\mu$
that relates the interface velocity and supercooling by (\ref{eq:Gibbs-Thomson})
has been adopted from \cite{Shibkov-interface_mobility} for a reference
supercooling of $1\text{ K}$. However, it has been known (and also
verified experimentally e.g. in \cite{Shibkov-interface_mobility})
that the relation between interface velocity and supercooling is not
generally linear.

\subsubsection*{Diffuse phase interface thickness and mesh resolution}

As discussed in detail in \cite{PF-Focusing-Latent-Heat}, the minimum
setting of the diffuse interface thickness parameter $\xi$ is restricted
by the resolution of the used numerical mesh. For a fixed $\xi>0$,
the solution of the phase-field problems is only an approximation
of the Stefan problem with surface tension \cite{Gurtin-Stefan_problem},
which would be attained as $\xi\to0$. Therefore, $\xi$ is also chosen
as small as possible. These considerations led us to setting $\xi$
proportional to the mesh cell size. Three different mesh resolutions
were used, as shown in Table \ref{tab:simulation-parameters-setup}.

\subsubsection*{Settings of $\gamma$ and $\zeta$}

The parameters $\gamma$ and $\zeta$ are involved in empirical formulas
(\ref{eq:explicit-p-dependence-on-T}) and (\ref{eq:water-indicator}),
respectively, proposed in this work for the model to follow the expected
qualitative behavior. As for phase transition ``rate'' $\gamma$
in the \textbf{Temp} model, the value $\gamma=2$ was used in most
simulations. With this value, the phase interface thickness in bulk
water during the cooling/heating phase is comparable to that obtained
by means of the phase field models. Increasing $\gamma$ further up
to $\gamma=10$ had negligible impact on the results (see Section
\ref{subsec:Macro-scale-results} below).

The reasoning behind the setting of $\zeta$ is as follows. Assume
that the volume occupied by liquid water and ice combined is defined
by
\begin{equation}
\Omega_{\text{W}}=\left\{ \left.\vec{x}\in\Omega\vphantom{\frac{1}{2}}\right|G\left(\vec{x}\right)<\frac{1}{2}\right\} .\label{eq:water-occupied-subdomain}
\end{equation}
Similarly, the volume occupied by ice is
\begin{equation}
\Omega_{\text{I}}\left(t\right)=\left\{ \left.\vec{x}\in\Omega\vphantom{\frac{1}{2}}\right|\phi\left(t,\vec{x}\right)>\frac{1}{2}\right\} .\label{eq:frozen-subdomain}
\end{equation}
As the temperature in the whole $\Omega$ drops significantly below
the freezing point of water at some $t>0$, it is natural to expect
that (essentially) all water eventually freezes to ice, i.e.,
\begin{equation}
\Omega_{\text{I}}\left(t\right)=\Omega_{\text{W}}.\label{eq:all-water-freezes}
\end{equation}
Then, the consequences for the setting of $\zeta$ depend on the model:
\begin{itemize}
\item In the \textbf{Temp} model, it follows from (\ref{eq:explicit-p-dependence-on-T})
that $\phi\approx I_{\text{W}}\left(G\right)$. The condition (\ref{eq:all-water-freezes})
can be rewritten as $\text{max}\left(0,1-\zeta G\right)>\frac{1}{2}\iff G<\frac{1}{2}$,
which is true for any $G\in\left[0,1\right]$ only if $\zeta=1$.
\item In both phase-field models, the reaction terms (\ref{eq:reaction-term-GradP})
and (\ref{eq:reaction-term-SigmaP1-P}) ensure that $\phi$ rises
from $0$ to $1$ together with the deepening supercooling at any
point where $I_{\text{W}}\left(G\right)>0$. Hence, we set $\zeta=2$
so that $I_{\text{W}}\left(G\right)$ is exactly zero outside of $\Omega_{\text{W}}$
and $\phi$ does not change there.
\end{itemize}

\subsubsection*{Nucleation site for the phase-field models}

The initial solid domain $\Omega_{\text{I,ini}}$ required by the
phase-field models (see Section \ref{subsec:Phase-Field-Approach})
should be small enough in order not to significantly influence the
simulation results, but still large enough so that it lasts until
the glass cap is cooled down below the freezing point of water. The
setting used for the simulations presented in this work was a thin
disk defined as

{\footnotesize
\begin{equation}
\Omega_{\text{I,ini}}=\left\{ \left.\vec{x}\in\Omega\vphantom{\left(x_{1}-\frac{L_{1}}{2}\right)^{2}}\right|0.87L_{3}<x_{3}<0.97L_{3}\wedge\left(x_{1}-\frac{L_{1}}{2}\right)^{2}+\left(x_{2}-\frac{L_{2}}{2}\right)^{2}<\left(\frac{L_{1}}{3}\right)^{2}\right\} .\label{eq:nucleation-site}
\end{equation}
}{\footnotesize\par}

\noindent Note that half of the disk virtually (according to the setting
of $\phi_{\text{ini}}$) intersects with the glass cap, where it remains
unchanged in the course of the whole simulation. However, thanks to
the form of (\ref{eq:material-properties-blending-formula}), it affects
neither the heat transfer nor the evolution of the phase field $\phi$.
\begin{table*}
\caption{\protect\label{tab:simulation-parameters-setup}Parameters setup for
the macro-scale simulations in Section \ref{subsec:Macro-scale-results}.
Parameters for all models are given, but each model utilizes only
a subset of them. All spatial dimensions are given relative to the
vessel height $L_{3}$, which is useful for the scaling introduced
later in Section \ref{subsec:Micro-scale-results}. The physical properties
of materials are set as in Table \ref{tab:Physical-quantities} except
for the value of $\alpha$.}

\centering{}%
\begin{tabular}{ccllc}
\toprule
Parameter & SI Unit & Description &  & \multicolumn{1}{c}{Value}\tabularnewline
\midrule
$L_{3}$ & $\text{m}$ & height of the vessel interior (dimension along $x_{3}$ axis) &  & $0.06$\tabularnewline
$L_{1}$ & $\text{m}$ & internal vessel width (dimension along $x_{1}$ axis) &  & $L_{3}/2$\tabularnewline
$L_{2}$ & $\text{m}$ & vessel depth (dimension along $x_{2}$ axis) &  & $L_{3}/2$\tabularnewline
$x_{3,\text{cap}}$ & $\text{m}$ & elevation of the bottom of the glass cap &  & \multicolumn{1}{c}{$\frac{11}{12}L_{3}$}\tabularnewline
$r$ & $\text{m}$ & glass spheres radius &  & $L_{3}/20$\tabularnewline
 &  &  &  & \tabularnewline
\multirow{3}{*}{$N_{3}$} & \multirow{3}{*}{1} & \multirow{3}{*}{number of mesh cells in the $x_{3}$ direction} & low resolution & $100$\tabularnewline
 &  &  & medium resolution & $200$\tabularnewline
 &  &  & high resolution & $400$\tabularnewline
$N_{1}$ & 1 & number of mesh cells in the $x_{1}$ direction &  & $N_{3}L_{1}/L_{3}$\tabularnewline
$N_{2}$ & 1 & number of mesh cells in the $x_{2}$ direction &  & $N_{3}L_{2}/L_{3}$\tabularnewline
 &  &  &  & \tabularnewline
\multirow{3}{*}{$\alpha$} & \multirow{3}{*}{$\text{m}^{-2}\,\text{s}$} & \multirow{3}{*}{coef. of attachment kinetics $\alpha=\frac{\Delta s}{\mu\sigma}$} & large $\sigma$ & $4.17\cdot10^{6}$\tabularnewline
 &  &  & small $\sigma$ & $1.25\cdot10^{7}$\tabularnewline
 &  &  & tiny $\sigma$ & $3.7\cdot10^{8}$\tabularnewline
$\xi_{G}$ & $\text{m}$ & thickness of the glass--water diffuse interface &  & $L_{3}/500$\tabularnewline
$\xi$ & $\text{m}$ & thickness of the ice--liquid diffuse interface &  & $L_{3}/N_{3}$\tabularnewline
$\varepsilon_{0}$ & 1 & \multirow{2}{*}{$\Sigma$ limiter parameters (eq. \ref{eq:Sigma-limiter}), set according
to \cite{PF-Focusing-Latent-Heat}} &  & $0.05$\tabularnewline
$\varepsilon_{1}$ & 1 &  &  & $0.2$\tabularnewline
$\gamma$ & $\text{K}^{-1}$ & rate of phase transition at the interface &  & \multicolumn{1}{c}{$2$ or $10$}\tabularnewline
$\zeta$ & 1 & water indicator function parameter &  & $1$ or $2$\tabularnewline
$T_{\text{ini}}$ & $\text{K}$ & initial temperature of the vessel content &  & $T^{*}+20$\tabularnewline
$T_{\text{freeze}}$ & $\text{K}$ & cooling gas temperature during the freezing phase &  & $T^{*}-25$\tabularnewline
$t_{\text{freeze}}$ & $\text{h}$ & duration of the freezing period &  & $5$\tabularnewline
$T_{\text{thaw}}$ & $\text{K}$ & heating gas temperature during the thawing phase &  & $T^{*}+20$\tabularnewline
$t_{\text{thaw}}$ & $\text{h}$ & duration of the thawing period &  & $5$\tabularnewline
\bottomrule
\end{tabular}
\end{table*}

\subsection{\protect\label{subsec:Macro-scale-results}Macro-scale parametric
study}

\begin{figure}
\begin{centering}
\includegraphics[width=0.98\figwidth]{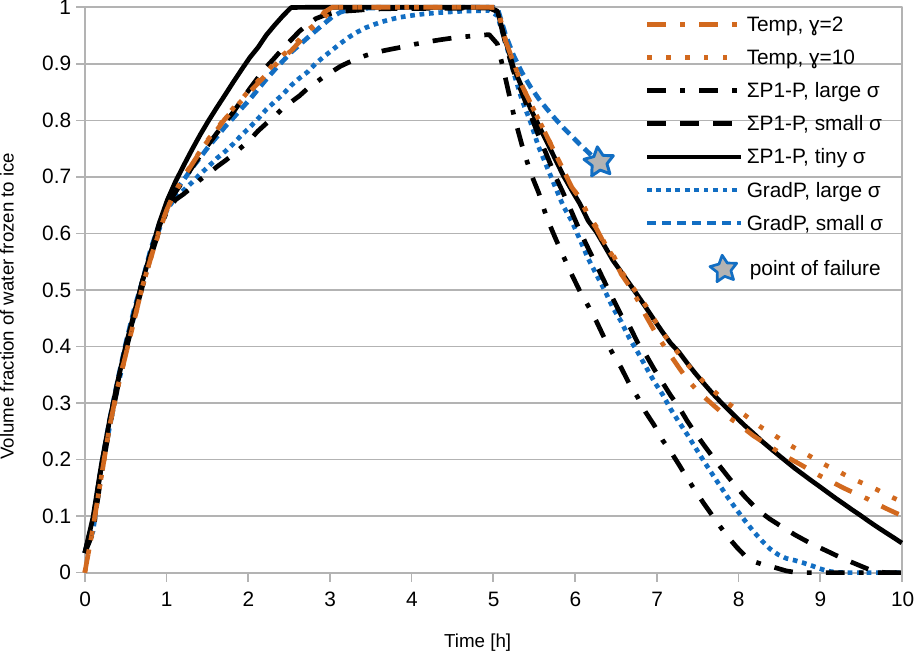}
\par\end{centering}
\caption{\protect\label{fig:model-comparison-ice-fraction-HR}Comparison of
the model results for medium mesh resolution ($N_{3}=200$). The evolution
of the ratio of the volume of ice (\ref{eq:frozen-subdomain}) and
the total water volume (\ref{eq:water-occupied-subdomain}) is presented.}
\end{figure}

First, we focus on model comparison based on a simple measure characterizing
the progression of freezing and thawing. In Figure \ref{fig:model-comparison-ice-fraction-HR},
the evolution of the volume fraction of ice $\frac{\left|\Omega_{\text{I}}\left(t\right)\right|}{\left|\Omega_{\text{W}}\right|}$,
i.e., the ratio of the volume of ice (\ref{eq:frozen-subdomain})
and the total water volume (\ref{eq:water-occupied-subdomain}) is
presented. Results obtained by all three proposed models (\textbf{GradP},
\textbf{$\mathbf{\Sigma}$P1-P}, and \textbf{Temp}), three values
of the surface tension $\sigma$ (``large'', ``small'', and ``tiny''
-- see Section \ref{subsec:Parameters-setup} and Table \ref{tab:simulation-parameters-setup}),
and two values of $\gamma$ are shown. The results can be interpreted
as follows.

\subsubsection*{Properties of the phase-field models}

The first observation is that the \textbf{GradP} model is unstable
and cannot deal with larger values of $\alpha$, which correspond
to smaller surface tension $\sigma$. In addition, \textbf{GradP}
seems to produce results with $\phi>1$, which is physically incorrect.
In contrast to that, the recently developed \textbf{$\mathbf{\Sigma}$P1-P}
model worked reliably in all situations. This behavior can be attributed
to the fact that the \textbf{GradP} model focuses the latent heat
interchange according to the actual shape of the diffuse phase interface,
which is distorted in the vicinity of the glass. In contrast to that,
the \textbf{$\mathbf{\Sigma}$P1-P} model distributes the latent heat
release and consumption across the diffuse interface according to
its theoretically predicted profile \cite{PF-Focusing-Latent-Heat},
which is not affected by the presence of glass or by the phase-field
diffusion strength being too low. Due to these findings, the \textbf{GradP}
model is not discussed in the subsequent evaluation.

\subsubsection*{Properties of the Temp model}

Figure \ref{fig:model-comparison-ice-fraction-HR} also shows the
sensitivity of the \textbf{Temp} model to the setting of $\gamma$.
In terms of the volume fraction of ice, the interface thickness setting
by means of different values of $\gamma$ has very little impact.
Without additional visual demonstration, let us note that the interface
thickness with $\phi$ given by the explicit relation (\ref{eq:explicit-p-dependence-on-T})
naturally increases as temperature levels out across $\Omega$. The
lower the temperature differences, the more pronounced is the impact
of $\gamma$ on the interface thickness. However, as $\phi$ is not
a primary variable in equations (\ref{eq:explicit-p-dependence-on-T})-(\ref{eq:combined-heat-xfer-phase-transition}),
the values of the temperature field are almost independent of $\gamma$.
Hence, as long as the frozen ice is distinguished from the liquid
by the relation (\ref{eq:frozen-subdomain}), both the volume and
the shape of the frozen subdomain $\Omega_{\text{I}}\left(t\right)$
are only weakly affected by the setting of $\gamma$.

\subsubsection*{Influence of the mesh resolution}

\begin{figure}
\begin{centering}
\includegraphics[width=0.98\figwidth]{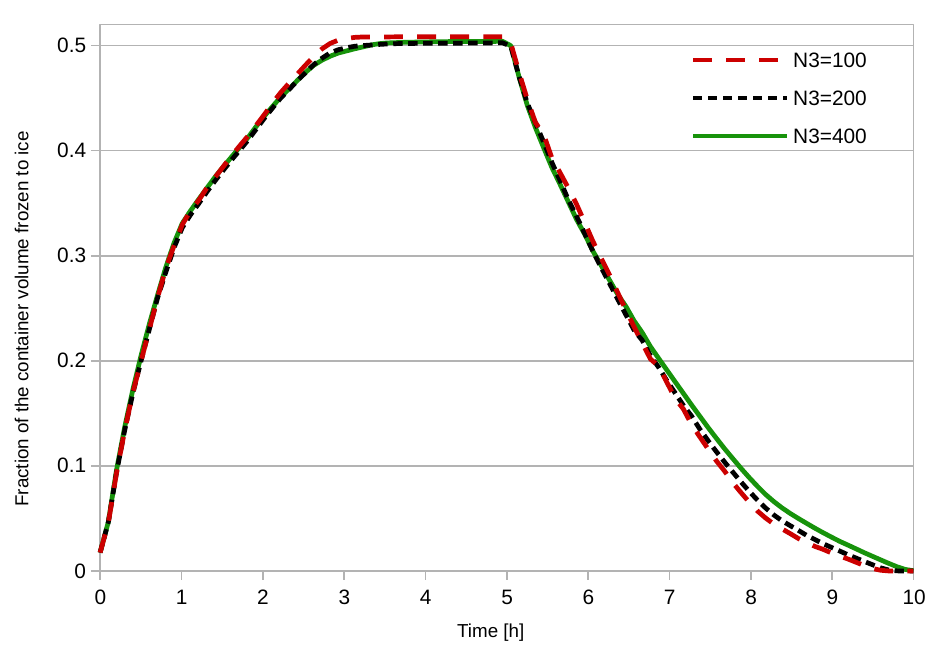}
\par\end{centering}
\caption{\protect\label{fig:SigmaP1-P_convergence-study}Dependence of the
course of freezing and thawing on mesh resolution. Results for the
\textbf{$\mathbf{\Sigma}$P1-P} model with small $\sigma$.}
\end{figure}

In addition to the simulations presented in Figure \ref{fig:model-comparison-ice-fraction-HR},
the computations were repeated on meshes with different resolutions:
low ($N_{3}=100$) and high ($N_{3}=400$). With the low mesh resolution,
\textbf{GradP} with small $\sigma$ did not work at all. For the \textbf{$\mathbf{\Sigma}$P1-P}
and \textbf{Temp} models, the curves of ice volume fraction $\frac{\left|\Omega_{\text{I}}\left(t\right)\right|}{\left|\Omega\right|}$
were compared. Note that this time, $\left|\Omega\right|$ (the volume
of the whole domain) was used in the denominator as the calculation
of $\left|\Omega_{\text{W}}\right|$ is resolution-dependent. For
the \textbf{Temp} model, the results on all three meshes are indistinguishable.
For the \textbf{$\mathbf{\Sigma}$P1-P} model, the comparison of the
results for different mesh resolutions is in Figure \ref{fig:SigmaP1-P_convergence-study},
which also indicates that the used meshes are fine enough for the
numerical solution to be considered accurate.

\subsubsection*{Influence of surface tension}

As for the effects of surface tension $\sigma$, the results (obtained
by the \textbf{$\mathbf{\Sigma}$P1-P} phase-field model) confirm
the expected behavior. For large $\sigma$, freezing occurs later
and thawing earlier compared to small $\sigma$, as the phase interface
hesitates to penetrate the cavities between the spheres and the ice
also thaws prematurely in the regions of high curvature. The same
conclusion applies to the comparison of small $\sigma$ with tiny
$\sigma$. With decreasing $\sigma$, the results of the \textbf{$\mathbf{\Sigma}$P1-P}
model resemble those of the \textbf{Temp} model, where surface tension
is absent. Note that the curves in Figure \ref{fig:model-comparison-ice-fraction-HR}
only start to differ as the freezing front reaches the porous structure.
Before that, all models predict identical behavior.

\subsubsection*{Three-dimensional visualizations}

In the series of Figures \ref{fig:snapshot07-freezing-above-bed}--\ref{fig:snapshot80-thawing-advances},
the time evolution of the freezing and thawing processes is demonstrated.
The results of the \textbf{$\mathbf{\Sigma}$P1-P} phase-field model
with both ``large'' and ``small'' $\sigma$ are compared to the
results of the \textbf{Temp} model. The results with ``tiny'' $\sigma$
are not included as they are visually very similar those of the \textbf{Temp}
model. Medium mesh resolution ($N_{3}=200$) was used for all three
simulations. Each figure provides an insight into the three-dimensional
geometry of the frozen subdomain $\Omega_{\text{I}}\left(t\right)$
by means of volume rendering of $\phi$ performed using the ParaView
\cite{ParaView}software package. In addition to that, a slice through
the temperature field is displayed.

\begin{figure}
\begin{centering}
\includegraphics[width=0.4\figwidth]{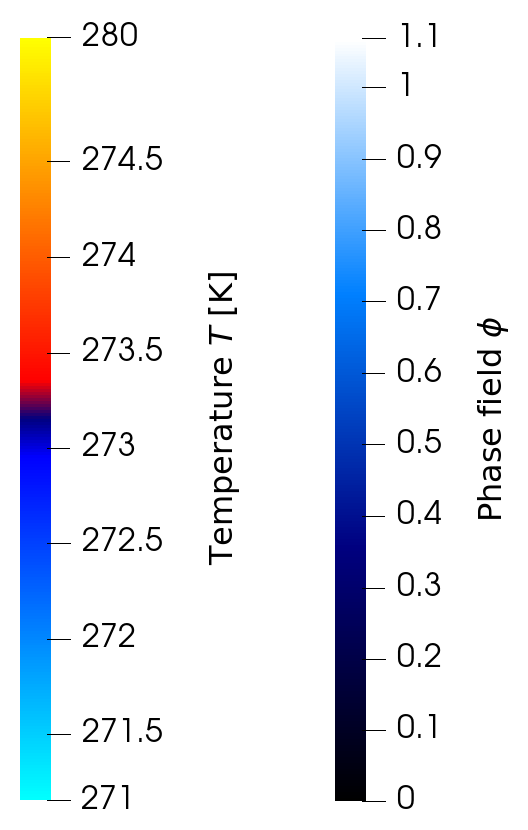}
\par\end{centering}
\caption{\protect\label{fig:Common-color-maps}Common color maps used in Figures
\ref{fig:snapshot07-freezing-above-bed}--\ref{fig:snapshot80-thawing-advances}.}
\end{figure}

\begin{figure*}
\begin{centering}
\subfloat[\textbf{Temp} model]{\begin{centering}
\includegraphics[width=0.32\textwidth]{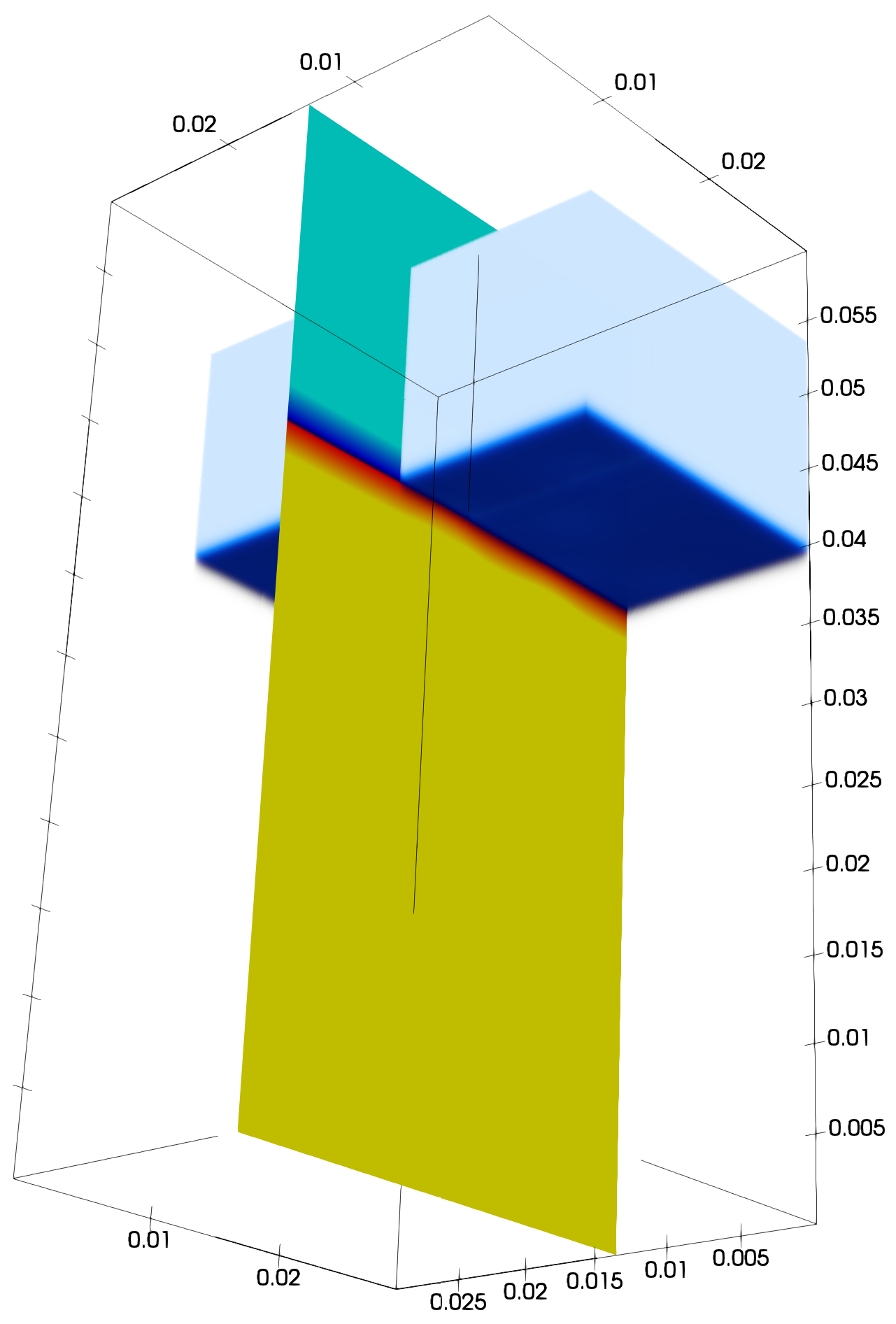}
\par\end{centering}

}\subfloat[\textbf{$\mathbf{\Sigma}$P1-P} model, small $\sigma$]{\begin{centering}
\includegraphics[width=0.32\textwidth]{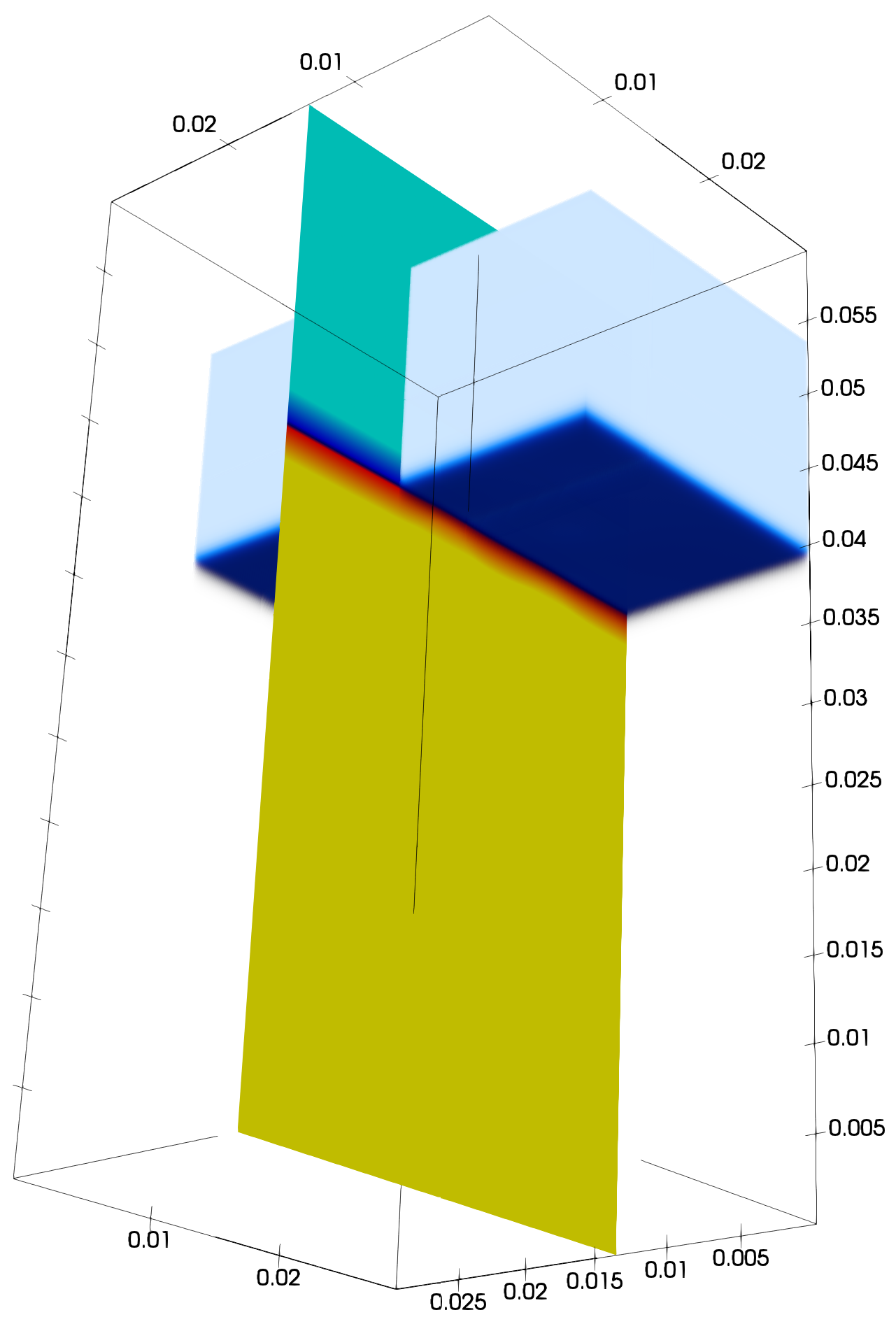}
\par\end{centering}
}\subfloat[\textbf{$\mathbf{\Sigma}$P1-P} model, large $\sigma$]{\begin{centering}
\includegraphics[width=0.32\textwidth]{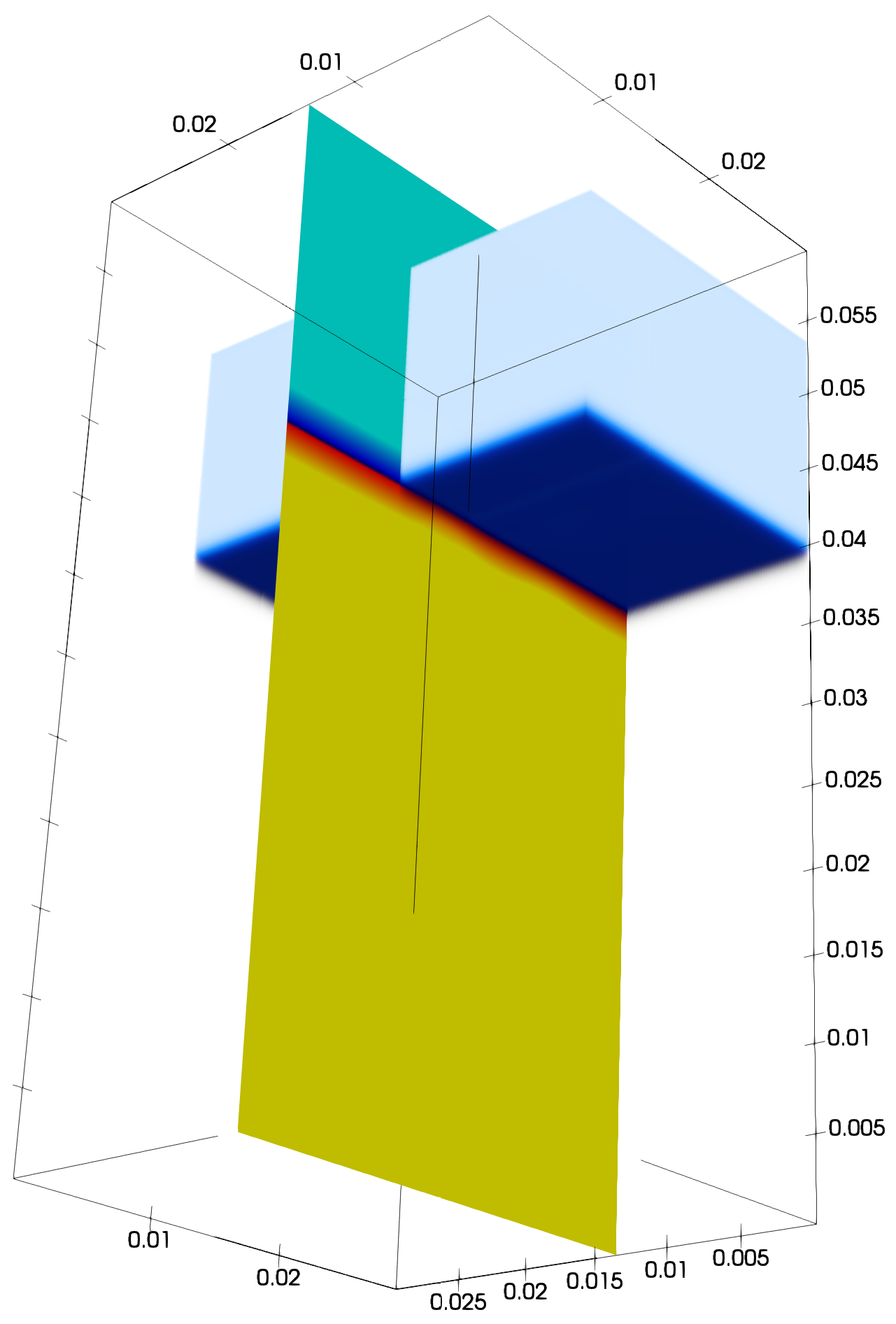}
\par\end{centering}
}
\par\end{centering}
\caption{\protect\label{fig:snapshot07-freezing-above-bed}3D visualization
of a subset of the frozen subdomain $\Omega_{\text{I}}\left(t\right)$
and a cut through the temperature field at time $t=42\text{ min}$.
The used color maps are in Figure \ref{fig:Common-color-maps}. The
freezing front propagates from above (from the bottom of the glass
cap) and it has yet to reach the top of the porous bed.}

\end{figure*}

\begin{figure*}
\begin{centering}
\subfloat[\textbf{Temp} model\label{fig:snapshot13-Temp-model}]{\begin{centering}
\includegraphics[width=0.32\textwidth]{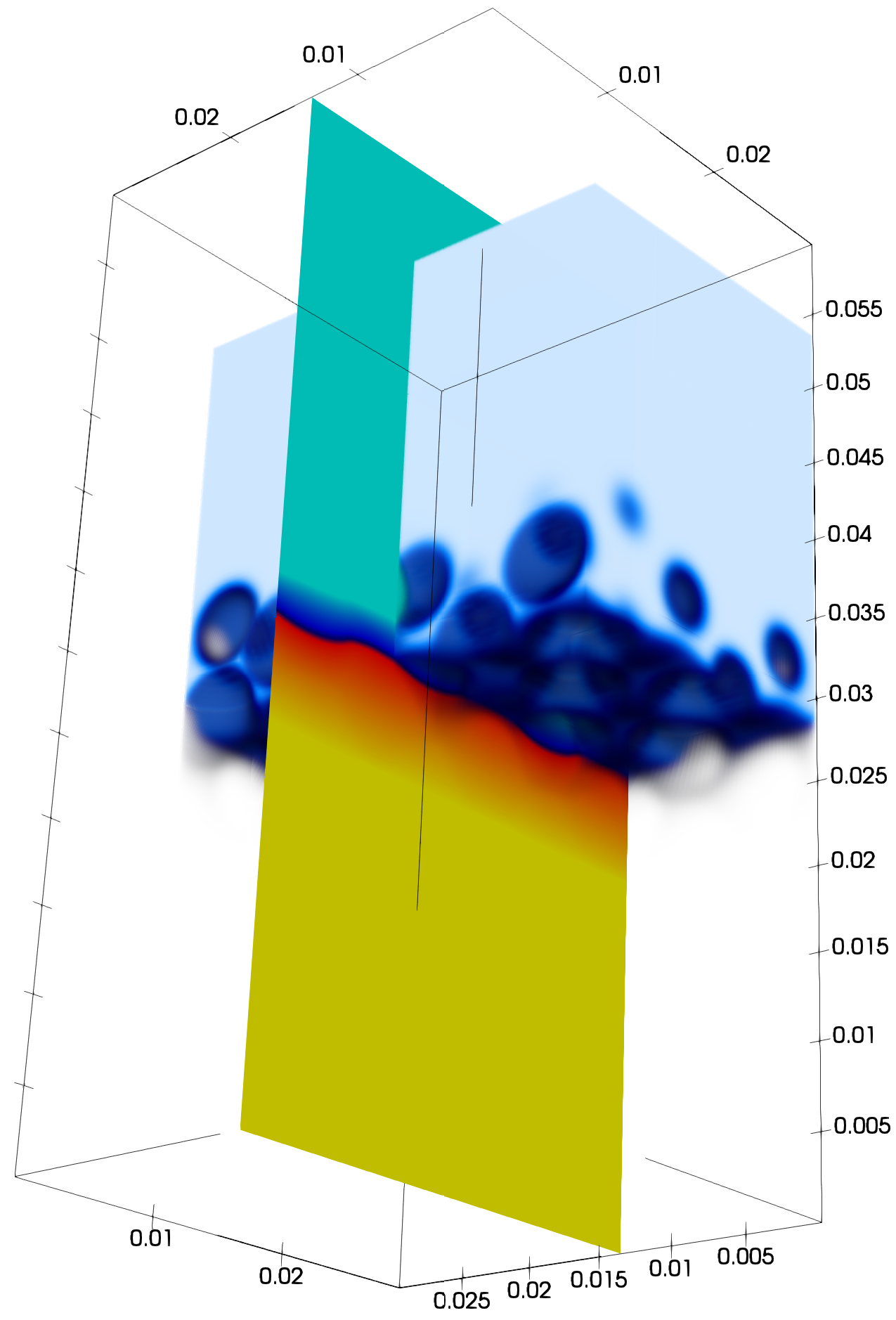}
\par\end{centering}
}\subfloat[\textbf{$\mathbf{\Sigma}$P1-P} model, small $\sigma$]{\begin{centering}
\includegraphics[width=0.32\textwidth]{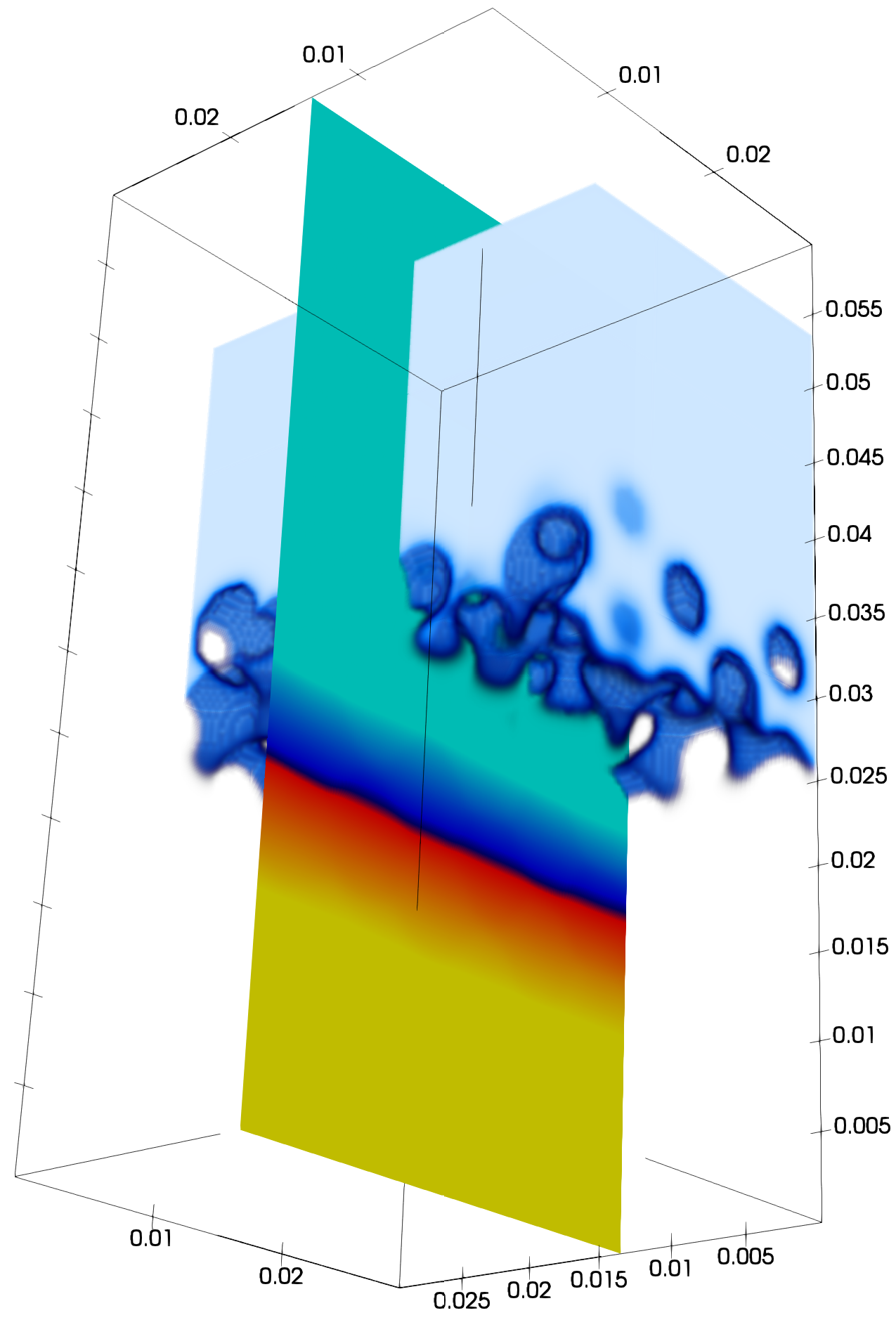}
\par\end{centering}
}\subfloat[\textbf{$\mathbf{\Sigma}$P1-P} model, large $\sigma$]{\begin{centering}
\includegraphics[width=0.32\textwidth]{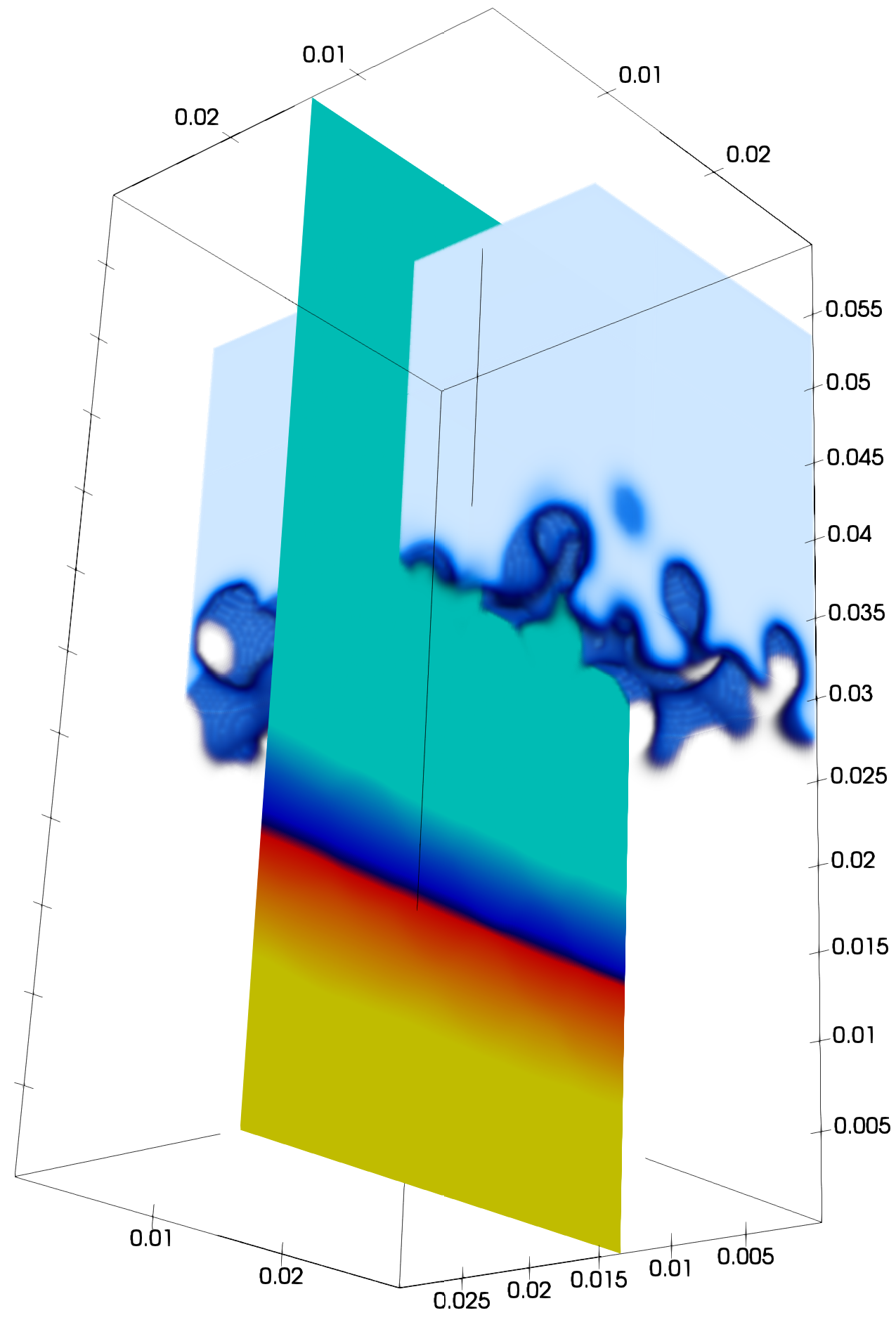}
\par\end{centering}
}
\par\end{centering}
\caption{\protect\label{fig:snapshot13-freezing-in-bed}3D visualization of
a subset of the frozen subdomain $\Omega_{\text{I}}\left(t\right)$
and a cut through the temperature field at time $t=1\text{ h }19\text{ min}$.
The used color maps are in Figure \ref{fig:Common-color-maps}. With
surface tension, the liquid close to the freezing front becomes supercooled.
This effect is even more pronounced for large $\sigma$, where, in
addition, the frozen region is remarkably less developed compared
to the other two results.}
\end{figure*}

\begin{figure*}
\begin{centering}
\subfloat[\textbf{Temp} model]{\begin{centering}
\includegraphics[width=0.32\textwidth]{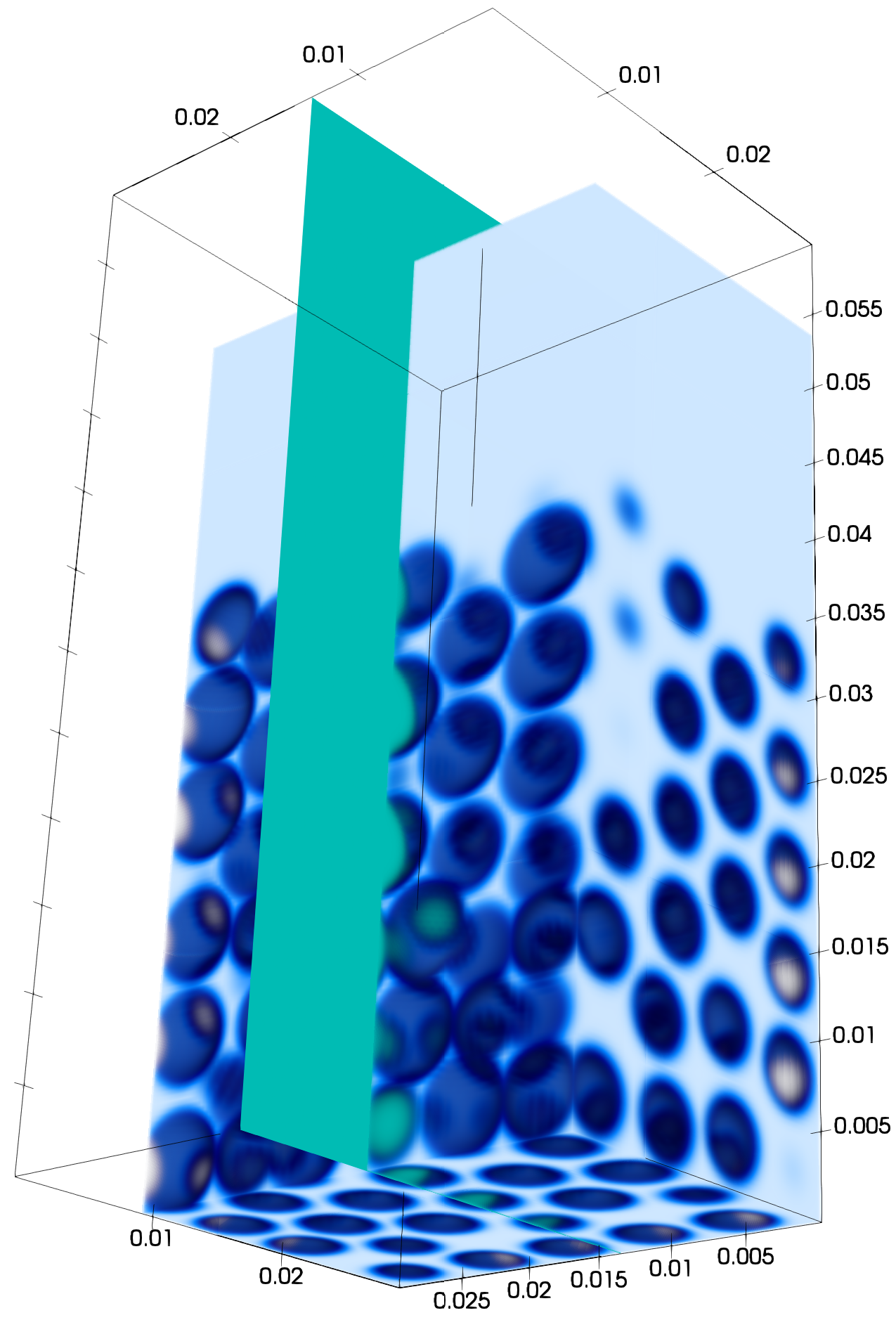}
\par\end{centering}
}\subfloat[\textbf{$\mathbf{\Sigma}$P1-P} model, small $\sigma$]{\begin{centering}
\includegraphics[width=0.32\textwidth]{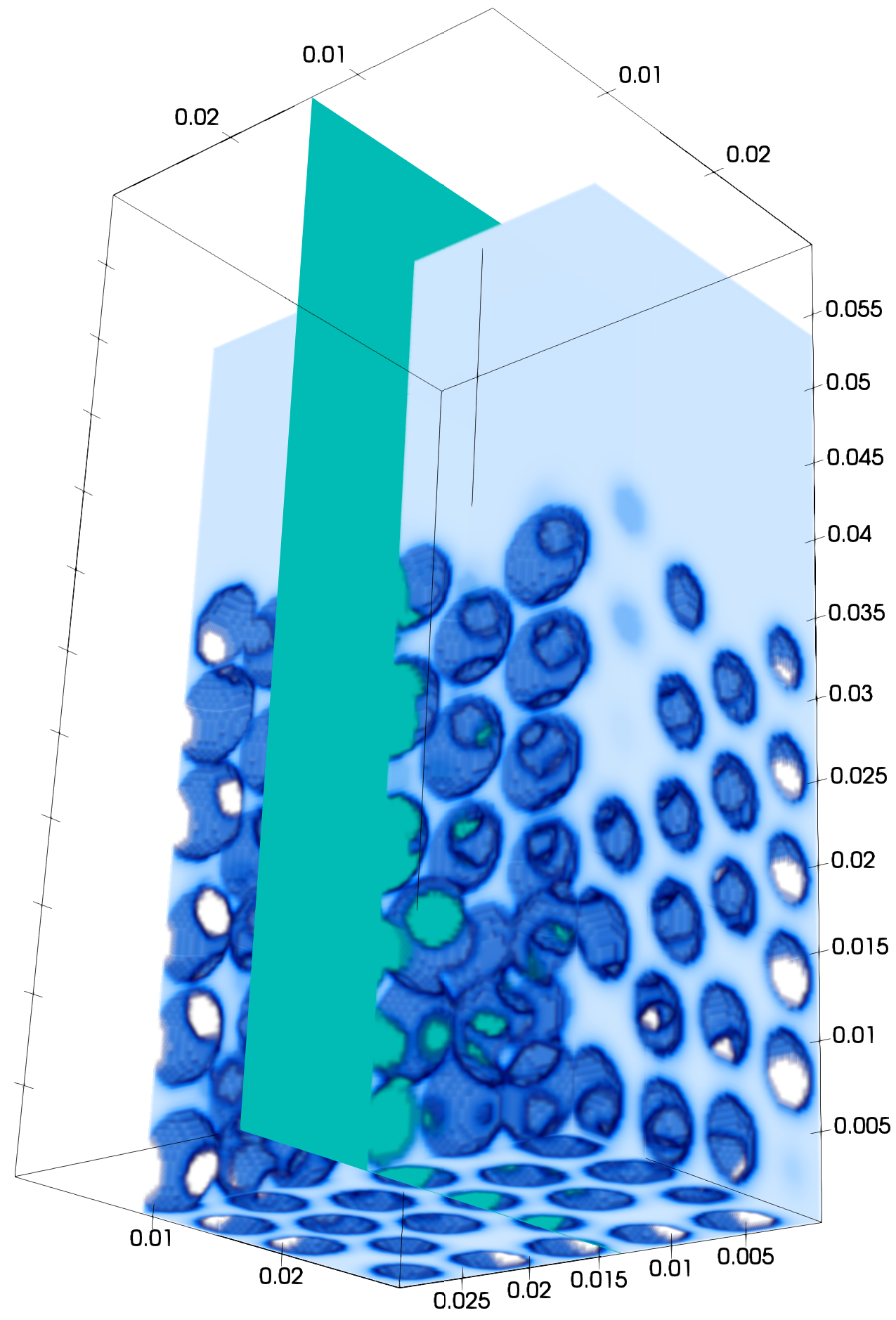}
\par\end{centering}
}\subfloat[\textbf{$\mathbf{\Sigma}$P1-P} model, large $\sigma$]{\begin{centering}
\includegraphics[width=0.32\textwidth]{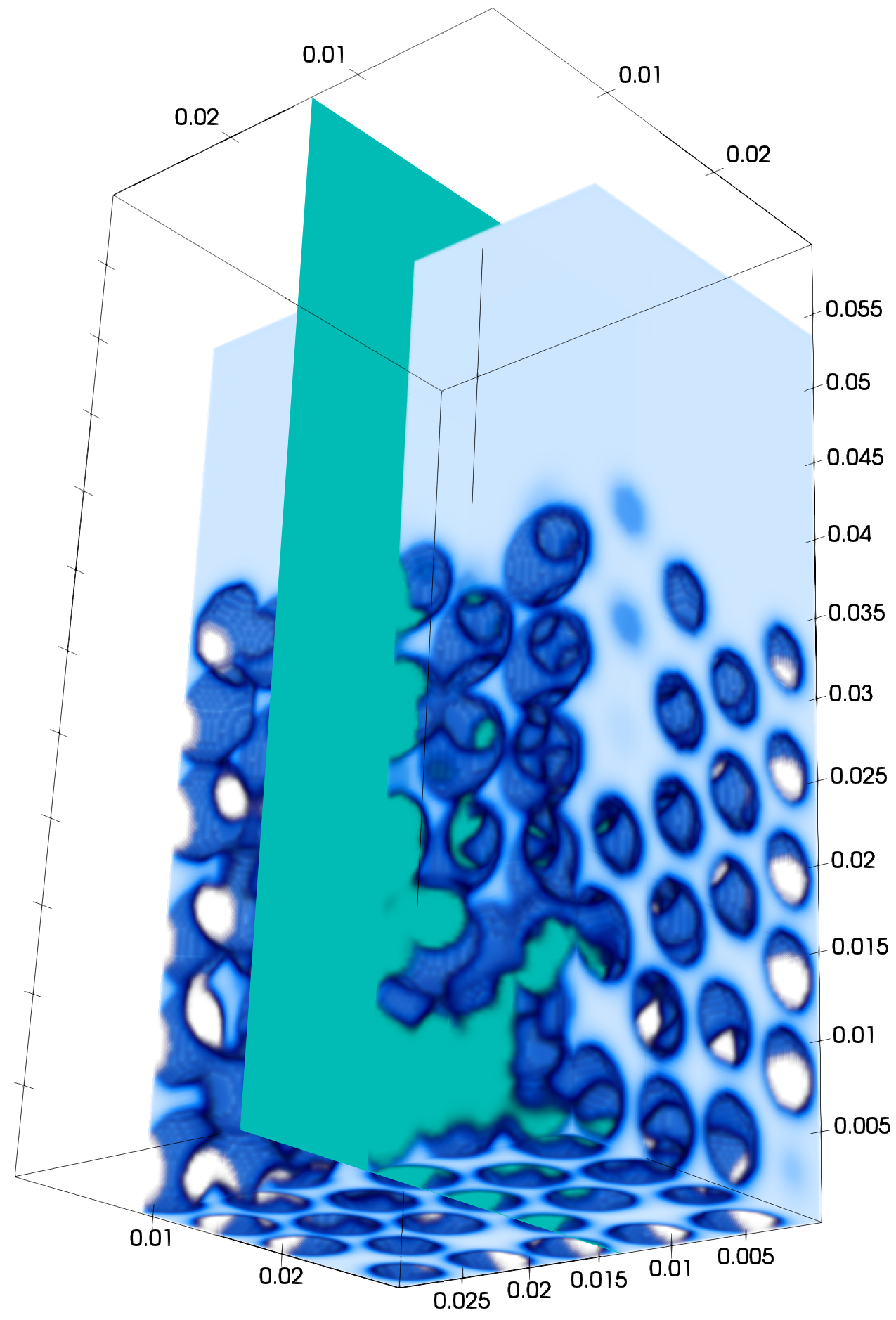}
\par\end{centering}
}
\par\end{centering}
\caption{\protect\label{fig:snapshot30-almost-frozen}3D visualization of a
subset of the frozen subdomain $\Omega_{\text{I}}\left(t\right)$
and a cut through the temperature field at time $t=3\text{ h }2\text{ min}$.
The used color maps are in Figure \ref{fig:Common-color-maps}. All
water is already frozen in the \textbf{Temp} model and in \textbf{$\mathbf{\Sigma}$P1-P}
with small $\sigma$. With large $\sigma$, some internal parts of
the porous structure still remain in liquid state.}
\end{figure*}

\begin{figure*}
\begin{centering}
\subfloat[\textbf{Temp} model]{\begin{centering}
\includegraphics[width=0.32\textwidth]{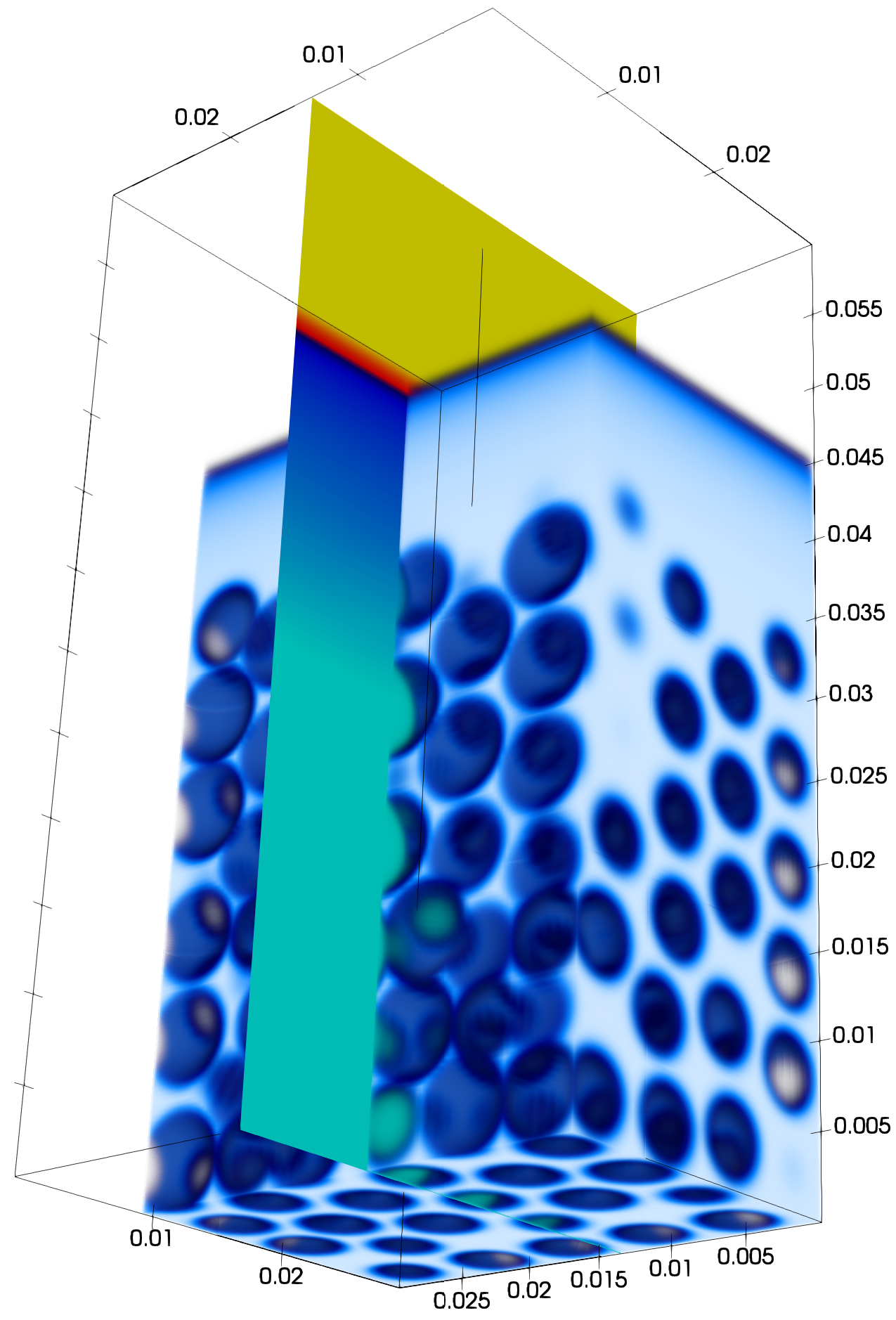}
\par\end{centering}
}\subfloat[\textbf{$\mathbf{\Sigma}$P1-P} model, small $\sigma$]{\begin{centering}
\includegraphics[width=0.32\textwidth]{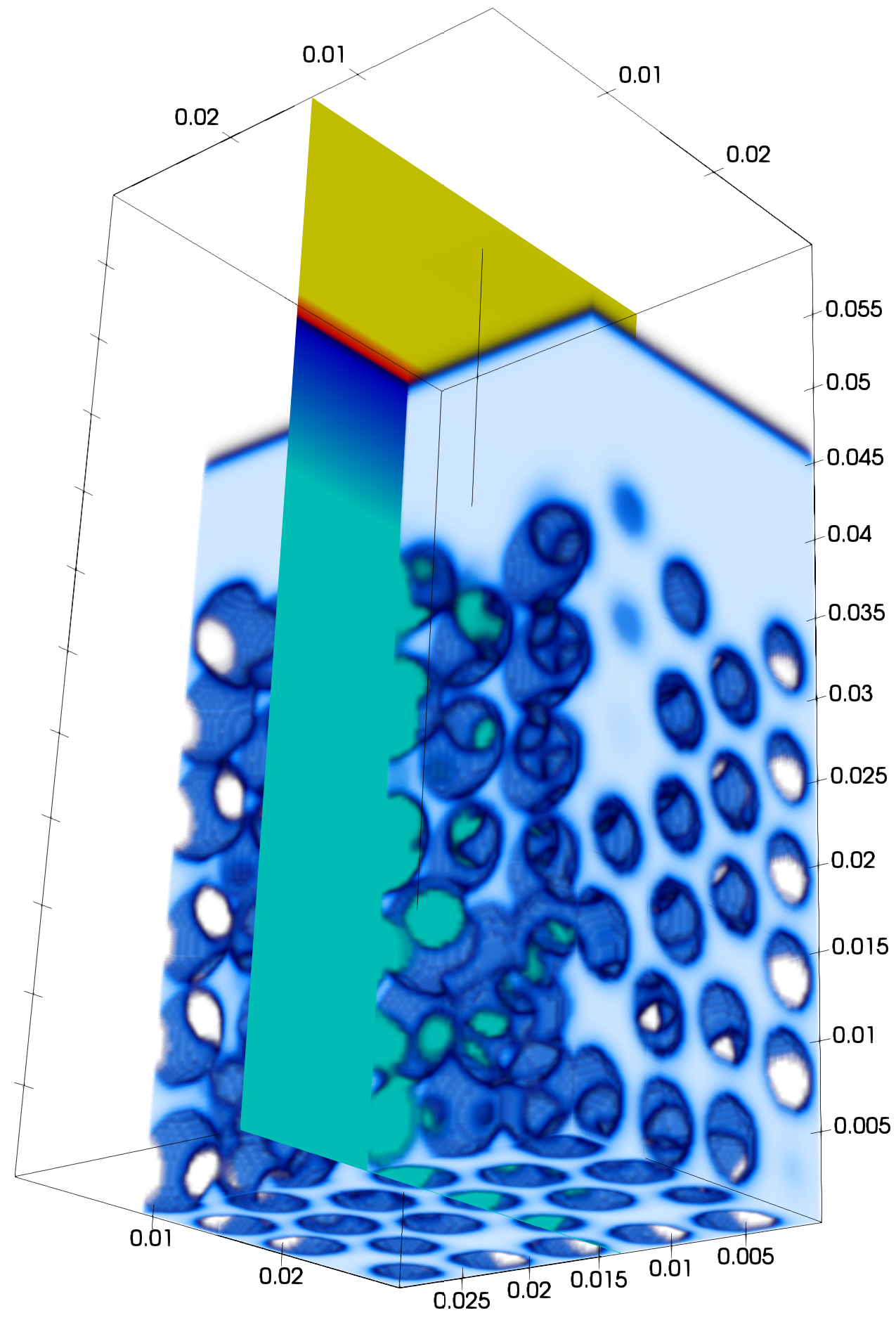}
\par\end{centering}
}\subfloat[\textbf{$\mathbf{\Sigma}$P1-P} model, large $\sigma$]{\begin{centering}
\includegraphics[width=0.32\textwidth]{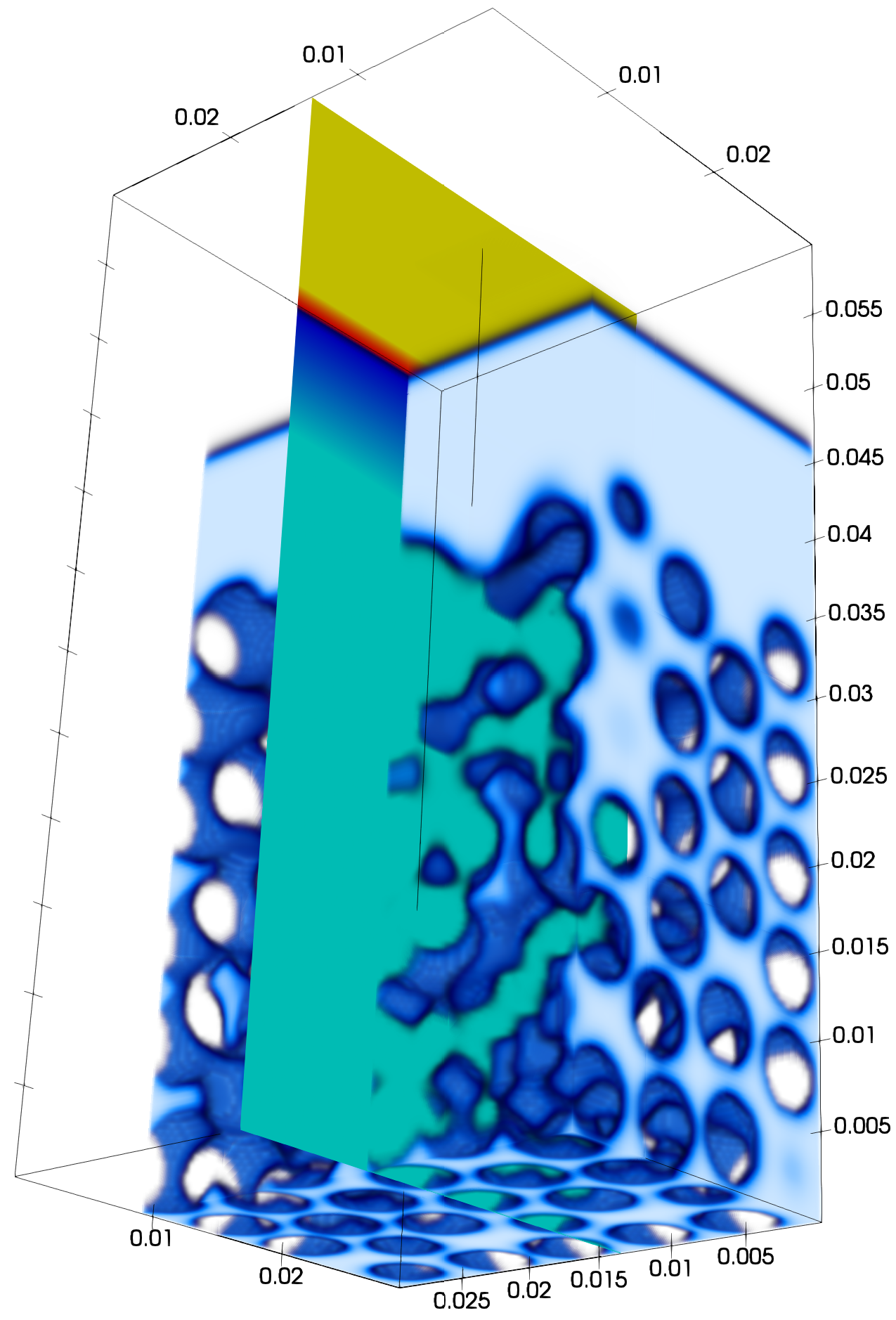}
\par\end{centering}
}
\par\end{centering}
\caption{\protect\label{fig:snapshot60-thawing-begins}3D visualization of
a subset of the frozen subdomain $\Omega_{\text{I}}\left(t\right)$
and a cut through the temperature field at time $t=6\text{ h }4\text{ min}$.
The used color maps are in Figure \ref{fig:Common-color-maps}. Thawing
propagates from the top. The top liquid-ice interface is aligned with
the the isosurface $T=T^{*}$. However, in \textbf{$\mathbf{\Sigma}$P1-P}
with large $\sigma$, thawing also takes place in the porous structure
below, which is still supercooled.}
\end{figure*}

\begin{figure*}
\begin{centering}
\subfloat[\textbf{Temp} model]{\begin{centering}
\includegraphics[width=0.32\textwidth]{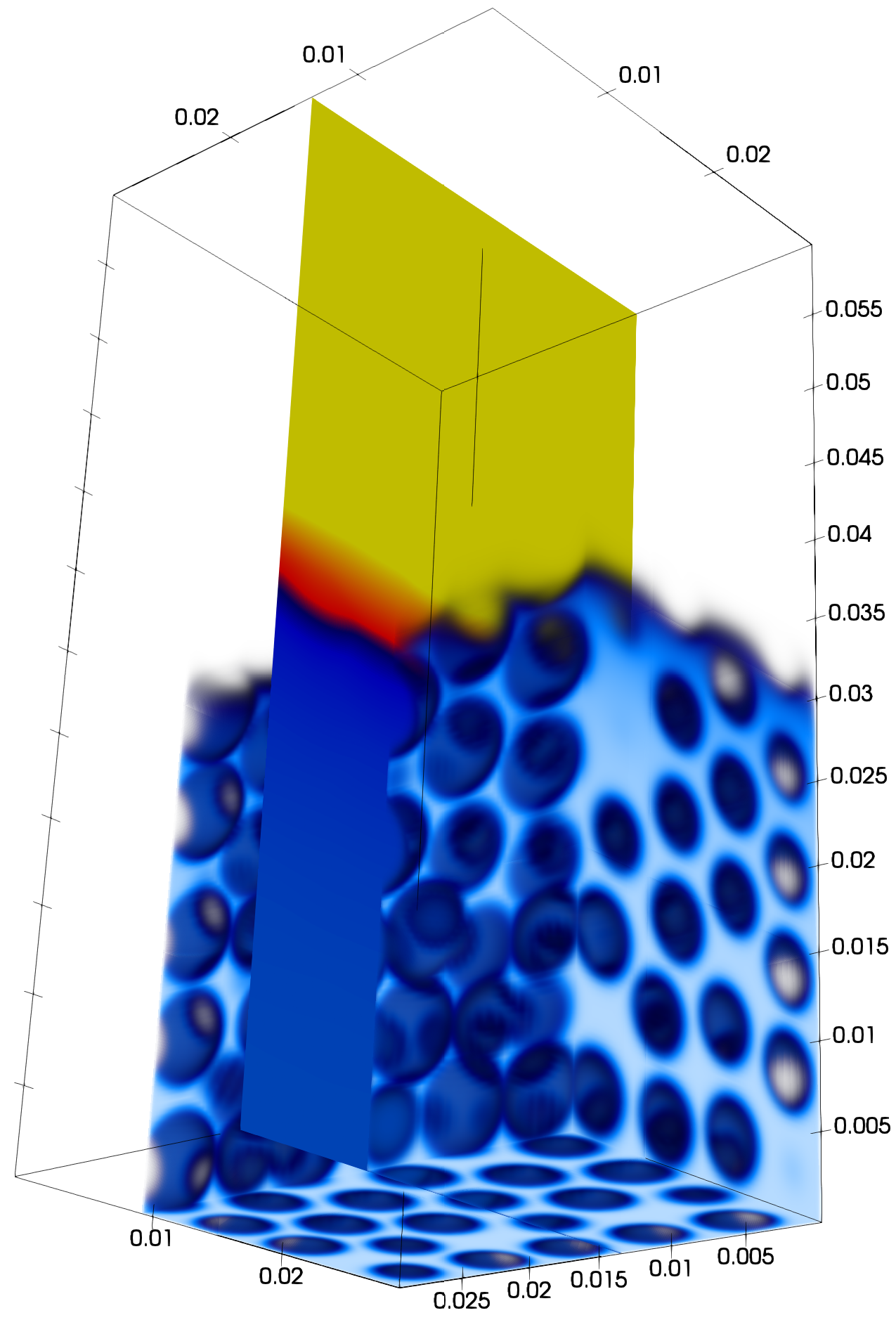}
\par\end{centering}
}\subfloat[\textbf{$\mathbf{\Sigma}$P1-P} model, small $\sigma$]{\begin{centering}
\includegraphics[width=0.32\textwidth]{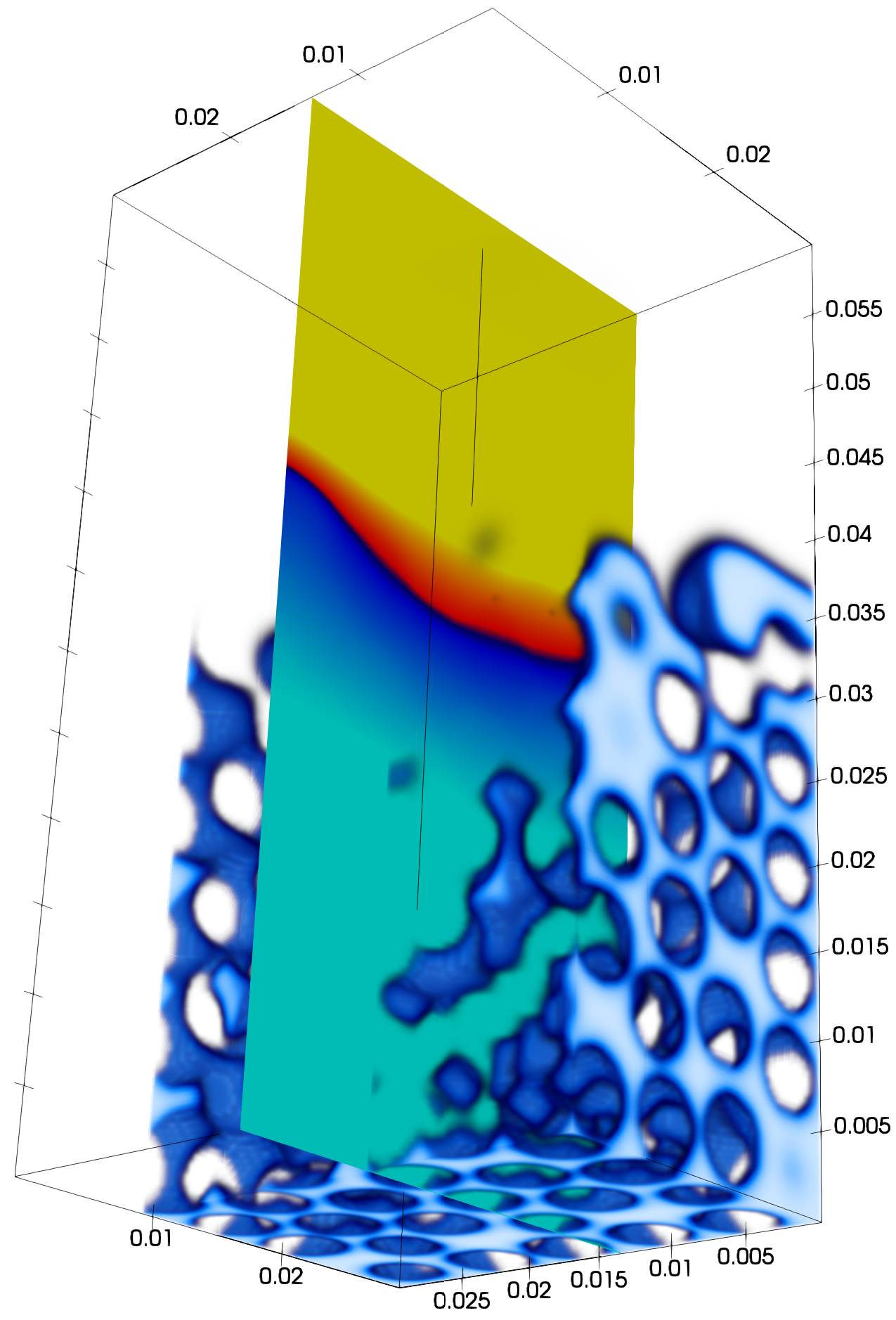}
\par\end{centering}
}\subfloat[\textbf{$\mathbf{\Sigma}$P1-P} model, large $\sigma$]{\begin{centering}
\includegraphics[width=0.32\textwidth]{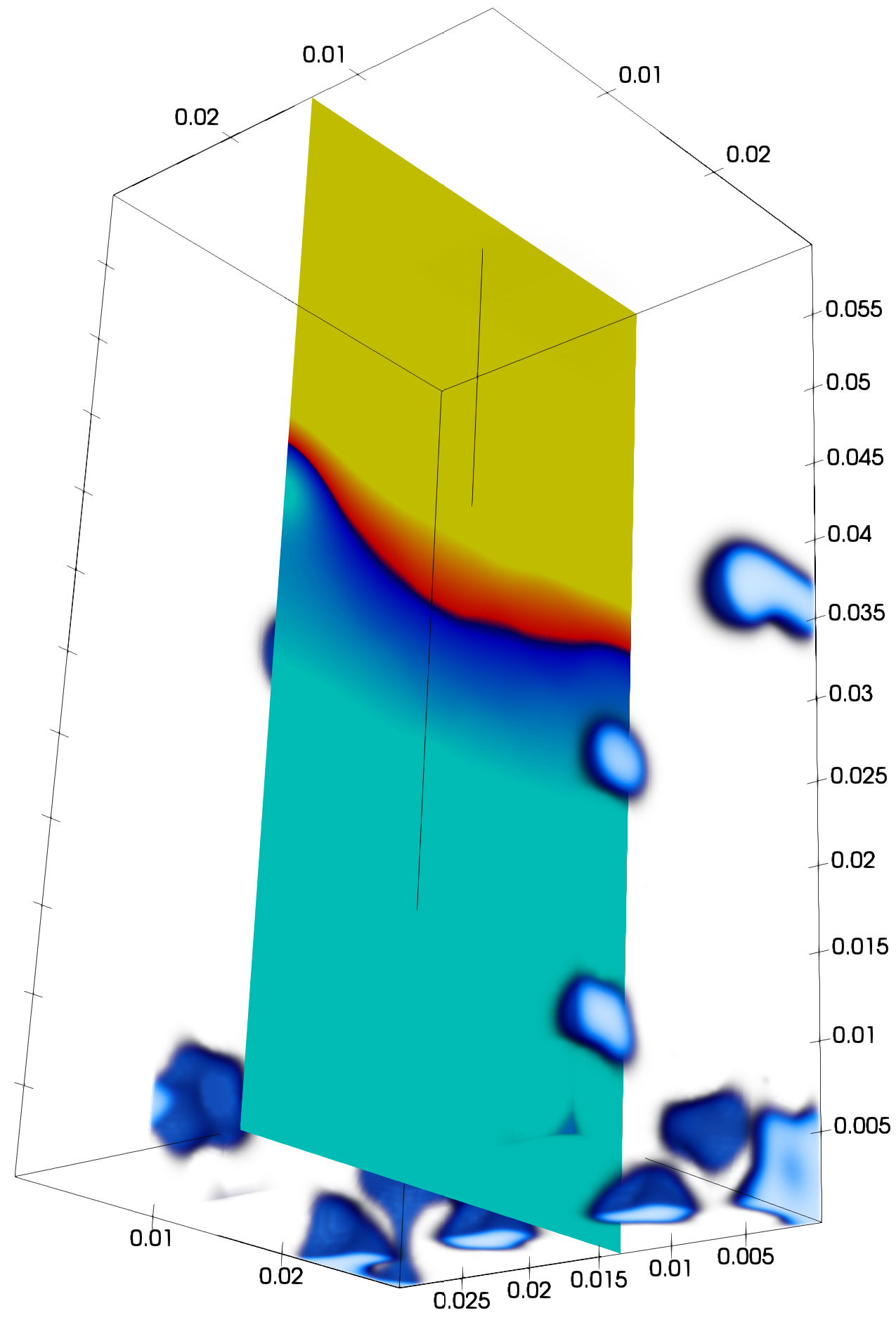}
\par\end{centering}
}
\par\end{centering}
\caption{\protect\label{fig:snapshot80-thawing-advances}3D visualization of
a subset of the frozen subdomain $\Omega_{\text{I}}\left(t\right)$
and a cut through the temperature field at time $t=8\text{ h }5\text{ min}$.
The used color maps are in Figure \ref{fig:Common-color-maps}. Striking
differences in the progression of thawing can be observed. In the
\textbf{Temp} model, interestingly, the temperature difference between
the thawing front and the bottom of the container is rather small.}
\end{figure*}

The color maps for $T$ and $\phi$ are identical across Figures \ref{fig:snapshot07-freezing-above-bed}--\ref{fig:snapshot80-thawing-advances}.
They are shown separately in Figure \ref{fig:Common-color-maps}.
The color map for the temperature $T$ is intentionally limited to
a narrow range to clearly visualize the regions with temperatures
above and below the freezing point $T^{*}$. The color map for the
phase field $\phi$ is useful in connection with the volume rendering
technique. The freezing front, where a diffuse interface between liquid
water and ice is formed, is rendered in darker colors. The boundaries
of the completely frozen regions are lighter.

The effects observable in the individual figures of the series are
as follows:
\begin{itemize}
\item In Figure \ref{fig:snapshot07-freezing-above-bed}, the freezing front
has yet to reach the porous bed below and all three results seem identical.
The boundary between ice and liquid water corresponds to the isosurface
$T=T^{*}$.
\item In Figure \ref{fig:snapshot13-freezing-in-bed}, ice starts to penetrate
the porous structure. While the geometry of the frozen region seems
similar in the \textbf{Temp} model and in \textbf{$\mathbf{\Sigma}$P1-P}
with small $\sigma$, the obvious difference consists in the position
of the temperature isosurface indicating the bulk water freezing point.
With surface tension, the liquid close to the freezing front becomes
supercooled. This effect is even more pronounced for large $\sigma$,
where, in addition, the frozen region is remarkably less developed
compared to the other two results.
\item In Figure \ref{fig:snapshot30-almost-frozen}, all water is already
frozen in the \textbf{Temp} model and in \textbf{$\mathbf{\Sigma}$P1-P}
with small $\sigma$. With large $\sigma$, some internal parts of
the porous structure still remain in liquid state.
\item In Figure \ref{fig:snapshot60-thawing-begins}, thawing propagates
from the top. The top liquid-ice interface is aligned with the the
isosurface $T=T^{*}$. However, in \textbf{$\mathbf{\Sigma}$P1-P}
with large $\sigma$, thawing also takes place in the porous structure
below, which is still supercooled.
\item Finally, in Figure \ref{fig:snapshot80-thawing-advances}, striking
differences in the progression of thawing can be observed. In the
\textbf{Temp} model, interestingly, the temperature difference between
the thawing front and the bottom of the container is rather small.
This is apparently due to heat conduction through the glass beads,
which have a larger thermal conductivity compared to water.
\end{itemize}
The quantitative behavior observed in the 3D visualizations is in
agreement with the conclusions based on Figure \ref{fig:model-comparison-ice-fraction-HR}.
In addition, the insight into ice formation and thawing in the three-dimensional
geometry of the pores provides better understanding of the differences
of the models and the effects of surface tension. Videos covering
the whole process of freezing and thawing are available online (refer
to the Data availability section below).


\subsection{\protect\label{subsec:Micro-scale-results}Unfrozen water content
determination at micro-scale}

The simulations in Section \ref{subsec:Macro-scale-results} used
values of $\sigma$ that were unrealistically large to demonstrate
the influence of surface tension on the qualitative aspects of the
results. Let us now attempt to investigate the phase-field approximation
of the Gibbs-Thomson relation quantitatively. For this purpose, a
transition to a much smaller scale is necessary.

The very fine particles of some porous materials create the conditions
for curvature-induced premelting, which is one of the causes of the
presence of unfrozen liquid at temperatures below the freezing point
of bulk water. There is a number of works that evaluate the unfrozen
water content in soils by means of experiments \cite{Zhang_et_al-Unfrozen_water_content_measurements_soil_2019,Watanabe-2017-unfrozen_water_measurements,Zhou_Zhou-Unfrozen_content_measurement-2013}
and calculations \cite{Wettlaufer_Worster-Premelting_review_2006,Nishimura-FEM_frozen_soil-2009,Rempel_Wettlaufer-Premelting_dynamics-2004}.
Such works take into account, e.g., the size distribution of the solid
particles, partial saturation of the pores by water \cite{Wei-Unsaturated_soils-2013},
the presence of impurities, and the chemical activity (salinity) of
the particles' surface. Saline solutions are also considered instead
of pure water \cite{Zhou_Zhou-Generalized-Clapeyron-unfrozen-2018}.
Therefore, data from these sources cannot be directly compared to
the results of our approach in its current form.

However, in \cite{Cahn_Dash_Fu-Premelting_monosized_powders-1992,Maruyama_Dash-Interfacial_melting_graphite_talc-1992,Dash-2006-premelting},
Dash et al. provide experimental and theoretical dependence of unfrozen
water content on supercooling for a situation well tailored to the
verification of our \textbf{$\mathbf{\Sigma}$P1-P} phase-field model.
In their research, pure water and special chemically inert powders
with narrow particle size distributions were used. In addition, careful
steps were taken to ensure that the porous medium was fully saturated.
The first material was polystyrene powder with average particle radius
$5\text{ }\mu\text{m}$. The second one was graphitized carbon black
with particle radius around $0.12\text{ }\mu\text{m}$.

\subsubsection*{Setup of micro-scale simulations}

For simulations, we essentially use the same geometrical setup as
in Section \ref{subsec:Macro-scale-results}, choosing the vertical
dimension of the vessel $L_{3}$ such that the particle radius $r$
corresponds to either of the above materials. The other length parameters
are calculated therefrom, as given in Table \ref{tab:simulation-parameters-setup}.
The value of $\alpha$ is realistic, i.e. set according to Table \ref{tab:Physical-quantities}.
As an equilibrium state under a constant supercooling $\Delta T=T^{*}-T_{\text{freeze}}$
is to be investigated, the approach of Section \ref{subsec:Micro-scale-approximation}
can be used. The ice proceeds from the nucleation site given by (\ref{eq:nucleation-site})
until the freezing process reaches a steady state.

We evaluated the dependence of the steady-state unfrozen water content
\begin{equation}
\text{UWC}=\left(1-\frac{\left|\tilde{\Omega}_{\text{I}}\left(t\right)\right|}{\left|\tilde{\Omega}_{\text{W}}\right|}\right)\cdot100\%,\label{eq:UWC-formula}
\end{equation}
on the supercooling $\Delta T$. In (\ref{eq:UWC-formula}), $\tilde{\Omega}_{\text{W}},\tilde{\Omega}_{\text{I}}\left(t\right)$
are analogous to (\ref{eq:water-occupied-subdomain}) and (\ref{eq:frozen-subdomain}),
respectively, except that only the bottom half of $\Omega$ fully
filled with particles is taken into account, representing a sample
taken from a larger container filled with the saturated porous medium.

A rough guess of the final time $t_{\text{freeze}}\approx\frac{L_{3}}{\mu\Delta T}$
corresponds to a moment of complete freezing if the ice front propagated
at the velocity $\mu\Delta T$ (velocity in the bulk water). This
value was adjusted experimentally for each case based on the real
delay of freezing in the porous region, ensuring that the freezing
process already reached the steady state (i.e., the value of $\text{UWC}$
did not change any further).

In \cite{Dash-2006-premelting}, the theoretical prediction of $\text{UWC}$
assumes that the spherical particles are arranged in the most compact
pattern possible, the face-centered cubic (fcc) lattice. For this
situation, the pores occupy around $26\%$ of the total volume \cite{Torquato-Random_close_packing-2000}.
However, due to friction, spherical particles usually settle in a
less organized structure known as random close packing, where the
pore (void) volume fraction can reach as high as $36\%$ \cite{Torquato-Random_close_packing-2000}.
To verify our settings, we evaluated the void fraction in the porous
structure used so far (Figure \ref{fig:spheres-and-cap}) to approx.
$29\%$, which is close to fcc. In addition, we reused our DEM tool
\cite{ALGORITMY2024-DEM_simulations-preprint} and by parameters tuning,
we obtained another porous structure with void fraction exactly $36\%$.
Both structures were used in the simulations.

\subsubsection*{Results and validation}

\begin{figure}
\begin{centering}
\includegraphics[width=0.98\figwidth]{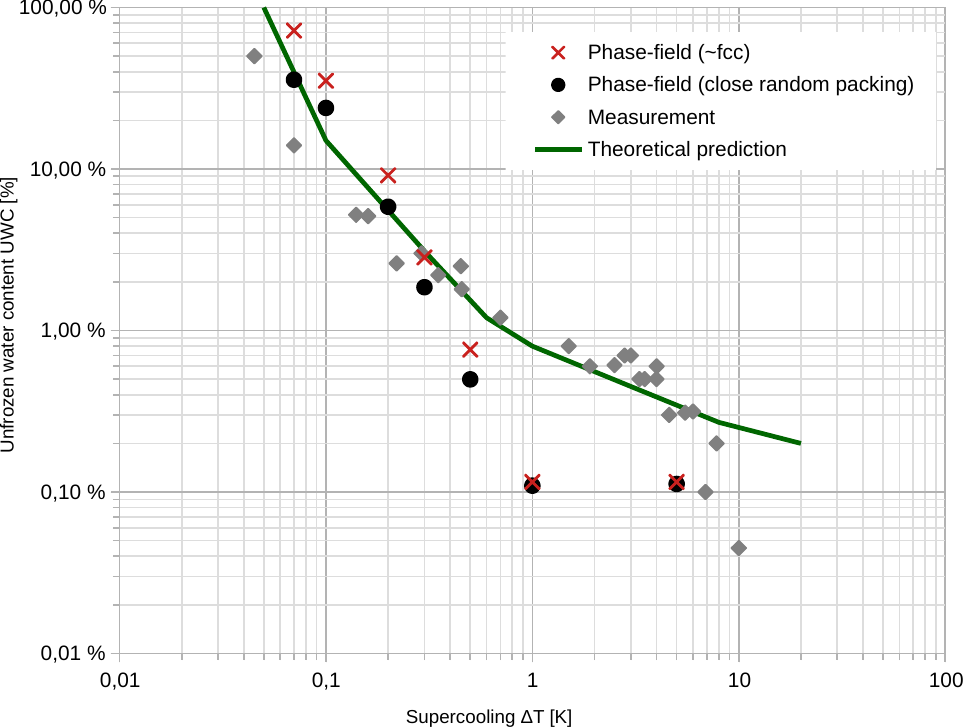}
\par\end{centering}
\caption{\protect\label{fig:UWC-polystyrene}The dependence of unfrozen water
content ($\text{UWC}$) on supercooling $\Delta T$ for a powder made
from polystyrene particles, fully saturated by water. Comparison of
predictions based on simulations with the \textbf{$\mathbf{\Sigma}$P1-P}
phase-field model and the results by Dash et al. \cite{Dash-2006-premelting}.
Two porous structures (nearly fcc and random close packing) were used
in the simulations.}

\end{figure}

\begin{figure}
\begin{centering}
\includegraphics[width=0.98\figwidth]{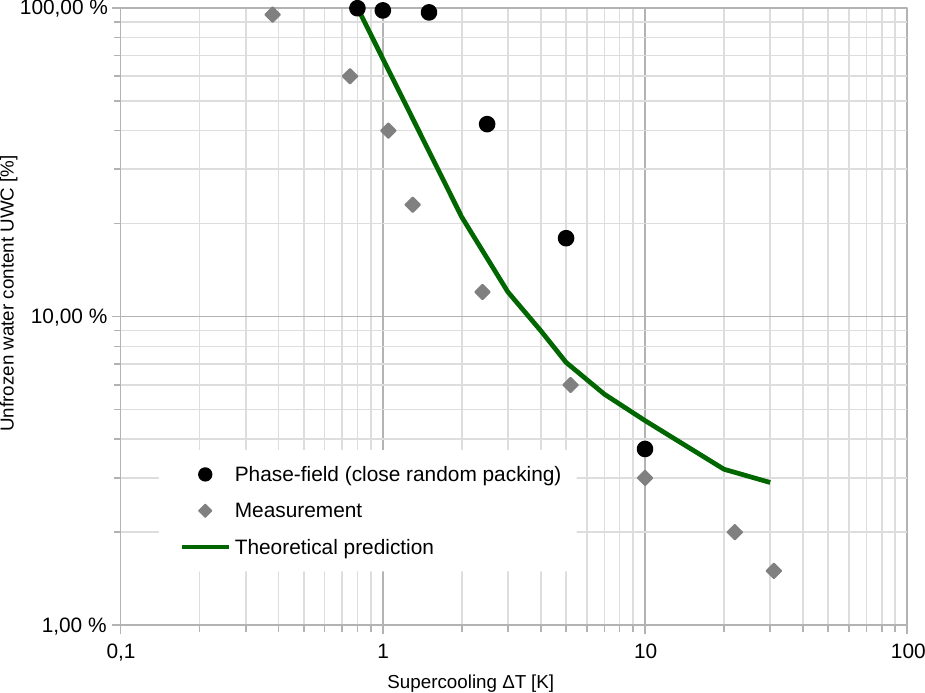}
\par\end{centering}
\caption{\protect\label{fig:UWC-graphitized_carbon_black}The dependence of
unfrozen water content ($\text{UWC}$) on supercooling $\Delta T$
for graphitized carbon black, fully saturated by water. Comparison
of predictions based on simulations with the \textbf{$\mathbf{\Sigma}$P1-P}
phase-field model and the results by Dash et al. \cite{Dash-2006-premelting}.
The porous structure with random close packing was used
in all simulations. Two more simulations for $\Delta T=10\text{ K}$
and $\Delta T=25\text{ K}$ yielded almost zero $\text{UWC}$, so
the data points are out of the range of the plot.}
\end{figure}

Figures \ref{fig:UWC-polystyrene}, \ref{fig:UWC-graphitized_carbon_black}
compare the predictions of $\text{UWC}$ obtained from the results
of phase-field simulations using the formula (\ref{eq:UWC-formula})
and the theoretical and experimental results by Dash et al. \cite{Cahn_Dash_Fu-Premelting_monosized_powders-1992,Maruyama_Dash-Interfacial_melting_graphite_talc-1992,Dash-2006-premelting}.
For polystyrene (Figure \ref{fig:UWC-polystyrene}), our model seems
to produce values in a good agreement with measurements, except for
the largest values of supercooling, where, supposedly, interfacial
premelting plays a more significant role. The organization of the
particles (and the associated void fraction) affects the results in
an expected manner, yielding lower values of $\text{UWC}$ for a larger
void fraction.

It is interesting to note that in the original paper \cite{Cahn_Dash_Fu-Premelting_monosized_powders-1992},
the radius of polystyrene particles was determined to be around $5\text{ }\mu\text{m}$
by means of electron microscopy. This is the value we used as well.
In a later text \cite{Dash-2006-premelting}, the authors correct
themselves and state that the radius must be just $1.5\text{ }\mu\text{m}$,
otherwise their theoretical prediction does not work. The size of
the particles is an essential parameter. Indeed, we also repeated
some of the simulations with particle size $1.5\text{ }\mu\text{m}$,
and the results were completely different. Hence, our results speak
in favor of the original measured value.

The agreement for graphitized carbon black (Figure \ref{fig:UWC-graphitized_carbon_black})
is a little less satisfactory. One of the reasons may be the fact
that graphitized carbon black contains polyhedral (i.e., non-spherical)
particles and the geometry of the void space is substantially different
from the one used in our simulations \cite{Graham-P33_carbon_black}.
Again, for the largest values of supercooling, our model predicts
almost zero $\text{UWC}$, whereas the experiments provide different
values.


\subsection{Computational costs\protect\label{subsec:Computational-costs}}

The simulations presented in Section \ref{subsec:Macro-scale-results}
were run on a high performance compute cluster, using MPI \cite{MPI-book-3_1}
to divide the computation into several processes (MPI ranks) launched
on one or more compute nodes. Each process utilized OpenMP \cite{OpenMP_Dagum}
to further divide the computation among multiple threads. Each compute
node was equipped with two 16-core AMD EPYC 7281@2.1GHz CPUs. The
nodes were interconnected by 100~Gbps OmniPath network, rendering
the speed of local and inter-node MPI communication comparable. The
software environment involved CentOS 7.9 Linux operating system, gcc
11.3 compiler with level 1 optimizations (-O1), and OpenMPI 2.1.5.

The computational costs of the simulations are summarized in Table
\ref{tab:Computational-costs}. Note again that for the phase-field
models, the interface thickness $\xi$ scales with mesh cell size,
which contributes to the increased number of time steps on finer meshes.
One can also observe that the simulations with the \textbf{GradP}
model were the most time demanding due to the numerical issues discussed
in Section \ref{subsec:Macro-scale-results}.

\begin{table*}
\caption{\protect\label{tab:Computational-costs}Representative computational
costs of the simulations for the individual models and different mesh
resolutions.}

\centering{}%
\begin{tabular}{ccccccccc}
\toprule
$N_{3}$ & Nodes  & Ranks & Threads & Total cores & \multicolumn{4}{c}{Computational time $[\text{HH}:\text{MM}]$}\tabularnewline
 &  &  &  &  & \textbf{$\mathbf{\Sigma}$P1-P}, large $\sigma$ & \textbf{$\mathbf{\Sigma}$P1-P}, small $\sigma$ & \textbf{GradP}, large $\sigma$ & \textbf{Temp}\tabularnewline
\midrule
100 & 1 & 4 & 8 & 32 & 01:11 & 01:15 & 02:42 & 00:23\tabularnewline
200 & 4 & 16 & 8 & 128 & 09:09 & 08:40 & 10:13 & 09:26\tabularnewline
400 & 12 & 24 & 16 & 384 &  & 90:31 &  & 74:17\tabularnewline
\bottomrule
\end{tabular}
\end{table*}

\section{Discussion and conclusion}

In this work, we have started the development of a framework for three-dimensional
simulation of freezing and thawing phenomena inside the complex geometry
of the pores, aimed at studying phase transition effects on a range
of spatial scales by means of different model variants. A small container
partially filled with spherical glass beads has been chosen as a model
of the porous medium at macro-scale. In this situation, the ice expansion,
fluid flow and mechanical interactions are not expected to be significant.
As the proposed mathematical model based on the phase-field approach
currently does not consider these phenomena, an already developed
parallel numerical solver could be easily adapted for its numerical
solution. The availability of experimental data may enable future
validation of the simulations.

The novel \textbf{$\mathbf{\Sigma}$P1-P} phase-field model introduced
in our recent work \cite{PF-Focusing-Latent-Heat} has been successfully
employed, leading to a numerically stable and efficient algorithm.
The used mesh resolution has been demonstrated to be sufficient in
terms of accuracy. The qualitative results of Section \ref{subsec:Macro-scale-results}
confirm that the phase-field model is able to capture the effects
of curvature-induced premelting, i.e., delayed freezing and premature
thawing inside the pores, which manifests itself as freezing point
depression on the macro-scale. As the surface tension is decreased,
the full phase-field model and the simplified \textbf{Temp} model
produce consistent results.

At the micro-scale, the phase-field approach allows to simulate equilibrium
states of a saturated porous medium at temperatures below the freezing
point of bulk water. The results allow to evaluate the unfrozen water
content. As long as the Gibbs-Thomson effect is the dominant mechanism
of premelting, a relatively good agreement with experimental and theoretical
data has been achieved for the case of ``monosized'' chemically
inert powders.

A number of directions of future research can be followed. The current
model deserves further studies, such as more thorough evaluation of
the sensitivity to numerical factors (mesh resolution, setting of
$\xi$) or sensitivity to the uncertainties in particle size and other
parameters. The DEM algorithms such as \cite{ALGORITMY2024-DEM_simulations-preprint}
easily allow to prepare porous beds consisting of particles with a
given particle size distribution, which could be utilized to investigate
unfrozen water content in real soils. This would in addition require
to simulate the mechanisms of interfacial premelting.

As for the validation against MRI data from dynamic freezing and thawing
experiments at the macro-scale, additional changes to model geometry
and overall setup may be necessary. For example, including the container
walls into the model may prove to be important with regard to heat
transfer. These changes are relatively easy to implement and some
steps in this direction have already been taken. If the additional
phenomena including ice expansion and mechanical interaction of ice
with the glass beads are found to be crucial for the correct results,
it may be appropriate to revise the phase-field approach currently
used to describe the distribution of glass inside the container. Alternatively,
some recent results using phase-field models in solid-fluid dynamics
problems \cite{Reder-PF-particulate_flows2021,Balashof-PF-solid_fluid_dynamics2023}
could be utilized.

\section*{Data availability}

The public GitHub repository at\\
\url{https://github.com/radixsorth/PorousFreezeThaw}\\
provides the following materials under MIT License:
\begin{itemize}
\item C++ source code of the hybrid parallel algorithm for simulations of
freezing and thawing in porous media
\item case setup files for the simulation cases presented in this article
\item 3D visualizations of the freezing and thawing simulation results
\end{itemize}
The source code of the DEM spherical particle dynamics simulator described
in \cite{ALGORITMY2024-DEM_simulations-preprint} with results and
visualizations is also included.

\section*{Declaration of competing interest}

The author declares that there are no known competing financial interests
or personal relationships that could have appeared to influence the
work reported in this paper.

\section*{CRediT authorship contribution statement}

\textbf{Pavel Strachota:} Conceptualization, Methodology, Software,
Validation, Formal analysis, Investigation, Data Curation, Writing
- Original Draft, Writing - Review \& Editing, Visualization.

%\begin{acknowledgements}

\section*{Acknowledgments}

This work is part of the project \emph{Multiphase flow, transport,
and structural changes related to water freezing and thawing in the
subsurface}, No. GA21-09093S of the Czech Science Foundation. Partial
support by the grant \emph{Modeling, prediction, and control of processes
in nature, industry, and medicine powered by high performance computing},
No. SGS23/188/OHK4/3T/14 of the Grant Agency of the Czech Technical
University in Prague, is gratefully acknowledged.

%\end{acknowledgements}

%\bibliographystyle{model1-num-names}
%\bibliography{References/references_FREEZING,References/references_MATH-PHYS,References/references_OPTIMIZATION,References/references_IT,References/publications,References/references_MMG,References/references_DEM}

\end{document}